\documentclass[12pt,a4paper,preprint,superscriptaddress]{revtex4-1}
\usepackage{subcaption}
\usepackage{float}
\usepackage{amsmath,amssymb}
\usepackage[utf8]{inputenc}
\usepackage{graphicx}
\usepackage[english]{babel}

\begin{document}

\title{Machine learning plasma-surface interface for coupling sputtering and gas-phase transport simulations}

\date{\today}
\author{Florian Krüger}
\affiliation{Institute of Theoretical Electrical Engineering, Ruhr University Bochum, 44801 Bochum, Germany}
\author{Tobias Gergs}
\affiliation{Institute of Theoretical Electrical Engineering, Ruhr University Bochum, 44801 Bochum, Germany}
\author{Jan Trieschmann}
\affiliation{Electrodynamics and Physical Electronics Group, Brandenburg University of Technology Cottbus--Senftenberg, Siemens-Halske-Ring 14, 03046 Cottbus, Germany}

\begin{abstract}
Thin film processing by means of sputter deposition inherently depends on the interaction of energetic particles with a target surface and the subsequent transport of film forming species through the plasma. The length and time scales of the underlying physical phenomena span orders of magnitudes. A theoretical description which bridges all time and length scales is not practically possible. A unified model approach which describes the dynamics of both the solid and the gas-phase, however, remains desired. In fact, advantage can be taken particularly from the well-separated time scales of the fundamental surface and plasma processes by evaluating both independently. Initially, surface properties may be \textit{a priori} calculated from a surface model and stored for a number of representative incident particle energy distribution functions. Subsequently, the surface data may be provided to gas-phase transport simulations via appropriate model interfaces (e.g., analytic expressions or look-up tables) and utilized to define insertion boundary conditions. During run-time evaluation, however, the maintained surface data may prove to be not sufficient (e.g., too narrow input data range). In this case, missing data may be obtained by interpolation (common), extrapolation (inaccurate), or be supplied on-demand by the surface model (computationally inefficient).
In this work, a potential alternative is established based on machine learning techniques using artificial neural networks. As a proof of concept, a multilayer perceptron network is trained and verified with sputtered particle distributions obtained from transport of ions in matter (TRIM) based simulations for Ar projectiles bombarding a Ti-Al composite. It is demonstrated that the trained network is able to predict the sputtered particle distributions for unknown, arbitrarily shaped incident ion energy distributions. It is consequently argued that the trained network may be readily used as a machine learning based model interface (e.g., by quasi-continuously sampling the desired sputtered particle distributions from the network), which is sufficiently accurate also in scenarios which have not been previously trained.
\end{abstract}

\maketitle

\newpage

\section{Introduction}
\label{sec:introduction}

Plasma based sputter processing is commonly used for the deposition of thin films (e.g., functional coatings for corrosion protection and wear-resistance applications or optical coatings) \cite{chapman_glow_1980,rossnagel_handbook_1990,lieberman_principles_2005,makabe_plasma_2006,martin_handbook_2010,sarakinos_high_2010}.
Within these processes, two characteristic features of technological plasmas are utilized: (i) The plasma supplies a continuous flux of ions directed toward the bounding surfaces. (ii) In front of these surfaces, the plasma establishes a thin space charge region (sheath) which is accompanied by a substantial local electric field and a drop in the corresponding potential (in particular at electrically driven electrodes). The amounts and the energy distributions of ion species from the plasma can be adjusted with the operating conditions (e.g., pressure and gas composition, voltage and waveform). \cite{chapman_glow_1980}

Sputtering itself is initiated prevalently by positively charged ions which are accelerated toward the electrode and act as projectiles, hitting the surface material with high kinetic energy. The momentum and kinetic energy of the projectiles is transferred to the surface (usually referred to as target) through a collision cascade. \cite{thompson_ii._1968,sigmund_theory_1969-1,sigmund_theory_1969,rossnagel_handbook_1990} The sputtering regime may be classified depending on projectile energy. With energies of a few hundred eV, technological plasmas typically yield binary collision conditions \cite{behrisch_sputtering_2007}. Resulting from this process, target material atoms are ejected into the plasma with a certain energetic and angular distribution (EAD) \cite{betz_energy_1994,stepanova_estimates_2001}.

The sputtered particles are transported through the plasma subject to collisions. These can be elastic collisions with neutral gas atoms/molecules or plasma ions causing background gas heating, \cite{hoffman_sputtering_1985} or inelastic collisions with energetic electrons causing excitation and ionization \cite{chapman_glow_1980,raizer_gas_1991,lieberman_principles_2005}.

When reaching a surface (in particular the substrate), particles may attach and form a thin solid film. This process is reliant on the EAD of impinging particles. Growth rate, crystalline structure and, consequently, mechanical and electrical properties of the formed films are just some of the affected properties \cite{martin_handbook_2010}. In particular the sputter yield, the ion-induced secondary electron emission coefficient, and the film's electrical conductivity have an inherent influence on the discharge operation \cite{depla_discharge_2006-1,depla_discharge_2006,depla_reactive_2008,depla_determination_2008}. A feedback loop with a strong bidirectional influence between plasma and surface establishes \cite{rossnagel_handbook_1990,berg_modeling_1987,berg_fundamental_2005,behrisch_sputtering_2007,martin_handbook_2010}.

Computer-based simulations play an integral role in understanding these interactions. Models may be formulated to describe the individual processes on their respective time and length scales. However, these scales combined span orders of magnitude from atomic to reactor scale and, hence, do not allow for a dynamical, simultaneous theoretical solution. As a remedy for this problem, advantage may be taken of the well-separated time and length scales by evaluating them independently. The models may be subsequently combined, bridging the different scales (e.g., using serial coupling) as suggested in the field of material science \cite{broughton_concurrent_1999,weinan_heterogeneous_2007}. This requires a consistent formulation of model interfaces, which is the focus of this work.

In terms of the individual models, technological plasmas are frequently analyzed using fluid models or particle in cell (PIC) simulations (depending on the operating regime), \cite{birdsall_plasma_1991,colonna_plasma_2016,van_dijk_plasma_2009} while the transport of sputtered particles is prevalently described via test particle methods (TPMs) \cite{somekh_thermalization_1984,turner_monte_1989,trieschmann_transport_2015}. For the specific case of ionized physical vapor deposition (PVD) such as high power impulse magnetron sputtering (HiPIMS), which is operated at low gas pressures below 1 Pa, self-consistent particle in cell/direct simulation Monte Carlo (PIC-DSMC) simulations may be required \cite{serikov_particle--cell_1999}. The mentioned gas-phase models resolve the plasma and particle dynamics on the reactor scale (i.e., process time and reactor dimensions). In contrast, the surface processes on the atomic scale are often studied by means of molecular dynamics (MD), kinetic Monte Carlo (kMC), or transport of ions in matter (TRIM) simulations \cite{biersack_monte_1980,eckstein_sputtering_1984,moller_tridyn_1984,voter_introduction_2007,graves_molecular_2009,neyts_molecular_2017}. The choice of the respective model depends severely on the physical process of concern and the investigated material system. A review of theoretical methods for modeling the plasma-solid interface with a focus on the solid dynamics is provided in \cite{bonitz_towards_2018}.

The data provided by one and required by another model may be exchanged between surface and gas-phase models via appropriate interfaces. For instance, the two models may be consistently coupled using the EADs of impinging and sputtered particles. This information exchange may be realized in terms of analytic expressions (e.g., Sigmund-Thompson distribution \cite{thompson_ii._1968,sigmund_theory_1969-1,sigmund_theory_1969}) or look-up tables (LUTs). Specifying accurate and computationally feasible interfaces in cases which involve only a few non-reactive species (e.g., argon and metal) is straightforward. In the simplest approach, the interfaces are reduced to integrated surface coefficients neglecting the particles' energy distributions (e.g., sputter yield, electron emission coefficient). Due to being rough estimates, this is not advised in the context of sputtering simulations even for the simplest systems. Taking into account the particle EADs (not integrated) provides a comparably simple yet more accurate description. However, complex surface and gas-phase systems (e.g., Ar, O$_2$, N$_2$ and Ti-Al-O-N) imply a higher order parameter space, complicating the interface models substantially \cite{berg_fundamental_2005,depla_reactive_2008}. As all combinations of species involved need to be parameterized, LUT approaches are rendered impractical.

In order to overcome the problems arising from the system complexity, we propose an alternative based on machine learning as a numerical plasma-surface interface. Artificial neural networks (ANNs) are trained using \textit{a priori} determined EADs of incident and sputtered particles. Ultimately, the aim is to predict sputtered particle EADs for arbitrary incident particle EADs, given three important features: i) reasonably abundant data is available from simulations and/or experiments, ii) generalization of the respective data is accomplished, iii) complex physics are preserved, which cannot be achieved by more simplified approaches. As a proof of concept, in this work fulfillment of these requirements is demonstrated using a multilayer perceptron (MLP) network. It is trained with and validated by sputtered particle EADs obtained from TRIDYN simulations \cite{moller_tridyn_1984} for the relative simple case of Ar projectiles bombarding a Ti-Al composite.

The manuscript is structured as follows: In Section~\ref{sec:scales} the simulation scenario and its inherent challenges are introduced. Subsequently, in Sections~\ref{sec:ANN} to \ref{sec:hyperparameters} our machine learning approach to the model interface is detailed. Thereafter, applicability of the proposed model is demonstrated and corresponding calculation results are presented in Section~\ref{sec:results}. Finally, the work is summarized and conclusions are drawn in Section~\ref{sec:conclusion}.

\section{Plasma-surface interface}

A plasma-surface interface for Ar projectiles incident on a Ti-Al composite target is developed in the following. The rather simple material system is chosen as an unsophisticated training and validation demonstrator. It allows for a detailed investigation of the prediction from incident to outgoing particle distributions using an MLP network. If desired, it is straightforward to extend the presented concept to different projectile species or surface composites (provided availability of training data).

\subsection{Physical scales}
\label{sec:scales}

The conceptual need for a plasma-surface model has been briefly introduced previously. To obtain a more extensive perspective, it is instructive to consider the fundamental physical scenario and its inherent time and length scales. Essentially three well-separated regimes have to be distinguished in technological sputter discharges: (i) heavy species gas-phase dynamics, which are on the order of macroscopic quantities; (ii) intermediate non-equilibrium electron gas-phase dynamics, which are dictated by the electrons' small mass; (iii) solid state surface dynamics, which inherently take place on the atomic level. Note that this classification ignores the electron dynamics in the solid, which are excluded for simplicity but may be taken into account if desired. The time and length scales are summarized in Figure~\ref{fig:scales}, while a detailed description is given below.

\begin{figure}[t]
\begin{center}
\includegraphics[width=8cm]{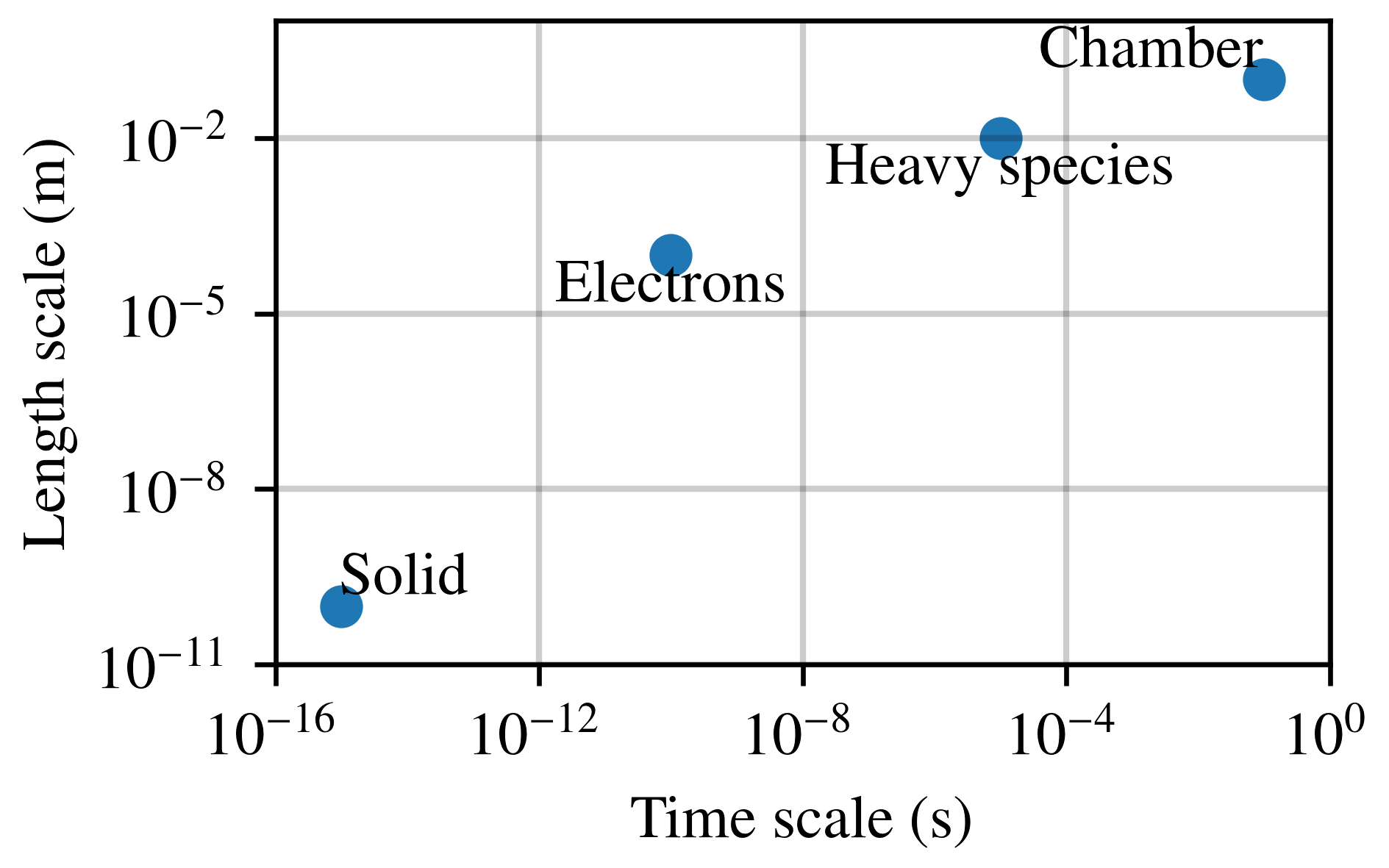}
\caption{Schematic of the physical time and length scales.}
\label{fig:scales}
\end{center}
\end{figure}

(i) Heavy species in the gas-phase are characterized by (close to) equilibrium gas dynamics. Correspondingly, their length scale is specified by the collision mean free path $\lambda_\text{c} \approx k_\text{B} T / (p \sigma) \approx 10^{-2}$~m (where $p \approx 0.7$~Pa is the gas pressure, $k_\text{B}$ is the Boltzmann constant, $T \approx 500$~K is the gas temperature, and $\sigma \approx 10^{-18}~\text{m}^2$ is the collision cross section). It is approximately on the order of the vacuum chamber dimensions $L \approx 10^{-2} \dots 10^{-1}$~m within PVD processes. In contrast, the time scale is dictated by the mean collision time, which is on the order of $\tau_\text{c} \approx \lambda_\text{c} / v_\text{th} \approx 10^{-5}$~s for heavy particle species (neutrals and ions) close to room temperature (i.e., with thermal velocity $v_\text{th}$). \cite{bird_molecular_1994}

(ii) Electron time and length scales are substantially different due to their small mass $m_\text{e} \approx 10^{-5} \, m_\text{Ar}$. The intrinsic length scale is the Debye screening length $\lambda_\text{D} = \sqrt{\epsilon_0 T_\text{e} / (e n_\text{e})} \approx 10^{-4}$~m, with elementary charge $e$, electron temperature $T_\text{e} \approx 3$~eV, electron density $n_\text{e} \approx 10^{16}~\text{m}^{-3}$, and vacuum permittivity $\epsilon_0$ \cite{chapman_glow_1980,lieberman_principles_2005}. In the low temperature plasma regime, the electron time scale is given by the inverse electron plasma frequency $\tau_\text{e} \approx \omega_{\text{pe}}^{-1} = 1 / \sqrt{e^2 n_e / (\epsilon_0 m_e)} \approx 10^{-10}$~s \cite{holt_foundations_1965,lieberman_principles_2005,wilczek_kinetic_2016}.

(iii) Solid state surface dynamics are governed by molecular interactions on the atomic level. In this regime, simulations must be able to resolve molecular oscillations in time and molecular bonds in space. Using the classical harmonic oscillator approximation, the vibrational frequency is given by $v_{\text{mo}} = \sqrt[]{K \mu^{-1}}$, with force constant $K$ and reduced mass $\mu$ \cite{larkin_infrared_2011}. With suitable magnitudes for $\mu \approx 10^{-27}$~kg and $K \approx 10^{-3} \, \textrm{N}\textrm{m}^{-1}$, a characteristic surface time scale $\tau_{\text{mo}} = v_{\text{mo}}^{-1} \approx 10^{-15}$~s is found. Regarding the length scale, the bond length $l$ in crystals in a face centered cubic (FCC) crystal system can be described by $l = \frac{a}{\sqrt{2}}$, with lattice constant $a$ \cite{callister_materials_2013}. For stoichiometric Ti-Al which crystallizes in FCC-type structure this yields a length scale on the order of $l \approx 10^{-10}$~m \cite{murray_calculation_1988}. The same order of magnitude is expected for amorphous Ti-Al.

Comparing the characteristic scales of the gas-phase and surface regime, it is found that $\tau_{\text{mo}} \ll \tau_{\text{e}} \ll \tau_\text{c}$ and $l \ll \lambda_\text{D} \ll \lambda_\text{c}$. The intrinsic time and length scales differ significantly, spanning about 10 orders of magnitude in time and 8 orders of magnitude in length. It is evident that defining time and length scales for a common simulation is not possible. The disparate models need to be considered separately and coupled appropriately.

The demonstration case selected in this work is concerned with the interaction of heavy particles (projectiles) with a solid surface. The link directed from gas-phase to solid surface is forthrightly established in terms of energy distributions of incident species. These are assumed to be available from a model (i.e., they are at this point assumed to be arbitrary, but physical). The goal is to establish a return link from the solid surface to the gas-phase based on an atomic scale surface model. It is infeasible to obtain the EADs of outgoing species for every possible incident species' energy distribution. Therefore, the surface model is initially solved for a representative set of \textit{a priori} specified cases. As detailed later, the simulation code TRIDYN is used for this purpose \cite{moller_tridyn_1984}. The obtained training data are then used to train an ANN model interface. The latter is finally intended to predict -- at run-time -- the EADs of outgoing species for arbitrary incident ones and, therewith, to specify the inflow of particles in the gas-phase model.

As the depicted demonstration case is concerned solely with heavy particle species, electron dynamics are ignored in this consideration. Moreover, also the individual heavy species gas-phase and solid surface models are detailed only as far as it relates to the plasma-surface interface to be established. A comprising and detailed treatment is out of the scope of this work.

\subsection{Data set generation and structure}
\label{sec:data}

Sufficiently sound and extensive sets of training data are indispensable for accurate ANN predictions. Fundamental methods like molecular dynamics (MD) may be utilized to provide these data sets at the expense of corresponding computing resources \cite{urbassek_molecular-dynamics_1997,graves_molecular_2009,brault_molecular_2015,neyts_molecular_2017}. To demonstrate applicability of the ANN model interface, however, transport of ions in matter (TRIM) based simulations are devised in this work, due to their significantly lower computational cost. General descriptions of the TRIM method and a discussion of detailed aspects can be found elsewhere \cite{biersack_monte_1980,eckstein_sputtering_1984,moller_tridyn_1984,behrisch_sputtering_2007,hofsass_simulation_2014}. Briefly, the interaction of projectiles incident on an amorphous solid compound and the dissipation of recoil cascades therein are simulated in binary collision approximation \cite{biersack_monte_1980}. Although strictly valid only at high projectile energies in the keV range, TRIM based simulations are widely applied also at lower energies \cite{behrisch_sputtering_2007}. For a sequence of impinging projectiles with a given energy distribution, these simulations provide the EADs of sputtered and reflected particles in terms of a sequence of outgoing particle properties (i.e., species, energy, directional cosines). TRIDYN \cite{moller_tridyn_1984} was used to generate training data of Ar impinging on a Ti-Al composite surface. While it allows for relatively low computation times compared to MD, it is nonetheless not suitable for run-time integration into gas-phase simulations as discussed in Section~\ref{sec:hyperparameters}.

Simulation results were obtained using the parameters listed in Table~\ref{table: global tridyn_parameters} and \ref{table:element specific tridyn_parameters}. The statistical quality of the training samples was intentionally limited to $N_\text{sp} = 10000$ incident projectiles. Using a larger number would improve the statistical quality of the results -- and is well within the capabilities of TRIDYN. However, since the presence of statistical noise allows for the investigation of important characteristics of the network, noisy distributions have been purposely chosen. A comparison with higher statistical quality results is presented subsequently.

\begin{table}[t]
\caption{Global TRIDYN parameters.}
\label{table: global tridyn_parameters}
\begin{center}
\begin{tabular}{ l   c   c }
\hline
parameter & symbol & value\\
\hline
no. of projectiles & $N_\text{sp}$ & $10^4$ \\
total fluence & $\Delta\Phi_\text{tot}$ & $ 3 \text{ \AA}^{-2}$ \\
pseudoatom weight & $N_0$ & $3000$\\ 
angle of radiation & $\theta$  & 0 \\
max. depth & $x_\text{max}$ &  1500 \text{\AA} \\
no. of depth intervals & $\Delta x $ & 500\\
\hline
\end{tabular}

\end{center}
\end{table}

\begin{table}[t]
\caption{Element specific TRIDYN parameters. $\varepsilon_{s,t}$ denotes the surface binding energy matrix elements, where index combinations $ s, t \,\epsilon \, \{ \text{Ar, Al, Ti} \}$ iterate over the individual contents of each target component, $ Qu_s $ is the respective initial atomic fraction, and $n_s$ the atomic density.}
\label{table:element specific tridyn_parameters}
\begin{center}

\begin{tabular}{ l   c    c   c   c   c}
\hline
element  & $\, \varepsilon_{s,\text{Ar}}$ (eV) &\,  $\varepsilon_{s,\text{Al}}$ (eV) &\,  $\varepsilon_{s,\text{Ti}}$ (eV) & \, $ Qu_s $ \,&\,  $n_s$ ($\text{\AA}^{-3}$) \\
\hline
Ar & 0 & 0 		& 0		& 0 	&2.49 \\
Al & 0 & 3.36	& 4.12	& 0.5	&6.02 \\
Ti & 0 & 4.12 	& 4.89	& 0.5	&5.58 \\
\hline

\end{tabular}
\end{center}
\end{table}

Each simulation is performed using a unique distribution of incident particles $f_{\text i}[l]$ with energy $E_l$ of the $l$-th bin. Because of the strong anisotropic acceleration of plasma ions in the sheath, ions are assumed to impact surface normal (i.e., polar angle $\theta = 0$ and no angular distribution). With regard to the training process, the energy distributions could be of arbitrary shape, but mono-energetic beams, discrete Gaussian and bimodal distributions have been chosen in order to represent physically relevant cases. The respective normalized, discrete functions are
\begin{eqnarray}
\label{eq:distribution1}
f_\text{i}[l] &=& \begin{cases}
1 & \text{if} \quad E = E_l \\
0 & \text{if} \quad E \neq E_l
\end{cases},
\\
\label{eq:distribution2}
f_\text{i}[l] &=& e^{-\frac{(E-E_l)^2}{\sigma^2}},
\\
\label{eq:distribution3}
f_\text{i}[l] &=& \begin{cases}
\sqrt{\frac{-1 + 2 \Delta_E}{\Delta_E^2 - (E - E_l)^2}} & \text{if} \quad \frac{E_l - E}{\Delta_E} 	\geq 1 \\
0 & \text{else}
\end{cases}.
\end{eqnarray}
Therein, $E_l$ denotes the energy of the $l$-th bin, $E$ is the distributions' median and mean energy, $\sigma^2$ is the variance, and $\Delta_E$ is the energy mode separation. \cite{zwillinger_standard_2002,lieberman_principles_2005}

TRIDYN evaluates the energy and the directional cosines of individual particles reflected by or sputtered of the target surface. For a sequence of incident projectiles, the sputtering events take place on a statistical basis. By creating histograms of the simulation output, each run gives the EADs $f_{\text{o},s}[m,n]$ with energy $E_m$ and surface normal cosine $c_n = \cos\vartheta_n$ of outgoing species $s$, as a function of the incident particle energy distribution $f_\text{i}[l]$.

To generate an entire data set, a variety of cases have been simulated, varying the mean energy $E$ from 0 to 1200 eV in steps of 8 eV for distributions \eqref{eq:distribution1}--\eqref{eq:distribution3}, respectively. For distribution \eqref{eq:distribution2}, the variance was set to $\sigma^2 =0.3$~eV, $\Delta_E = 30$~eV  was chosen for distribution \eqref{eq:distribution3}. In general, bigger training sets provide trained ANNs with better predictive qualities. However, in this work the number of sets was intentionally limited to 439 (i.e., 150 per distribution \eqref{eq:distribution1}--\eqref{eq:distribution3} with 11 omitted at the outskirts, due to the width of the peaks) to provide a proof of concept also for computationally limited cases (e.g., MD).

\subsection{Artificial neural network}
\label{sec:ANN}

An ANN is a collection of interconnected processing units called neurons or nodes. ANNs can be used to approximate the relationship between given input and output vectors $x_j$ and $y_k$, generally without being programmed with any task-specific rules \cite{bhadeshia_neural_1999,bishop_neural_1996,haykin_neural_2008}. At each node $k$, all inputs $x_j$ are multiplied with a weight $w_{kj}$. A constant bias $b_k$ may be introduced additionally. A sum is taken over all $d$ input nodes and the result is passed to a nonlinear activation function $\phi$, which gives the output $y_k = \phi\left(b_k + \sum_{j=0}^{d}{w_{kj} x_j} \right)$ of node $k$.

\begin{figure}[t]
\begin{center}
\includegraphics[width=8cm]{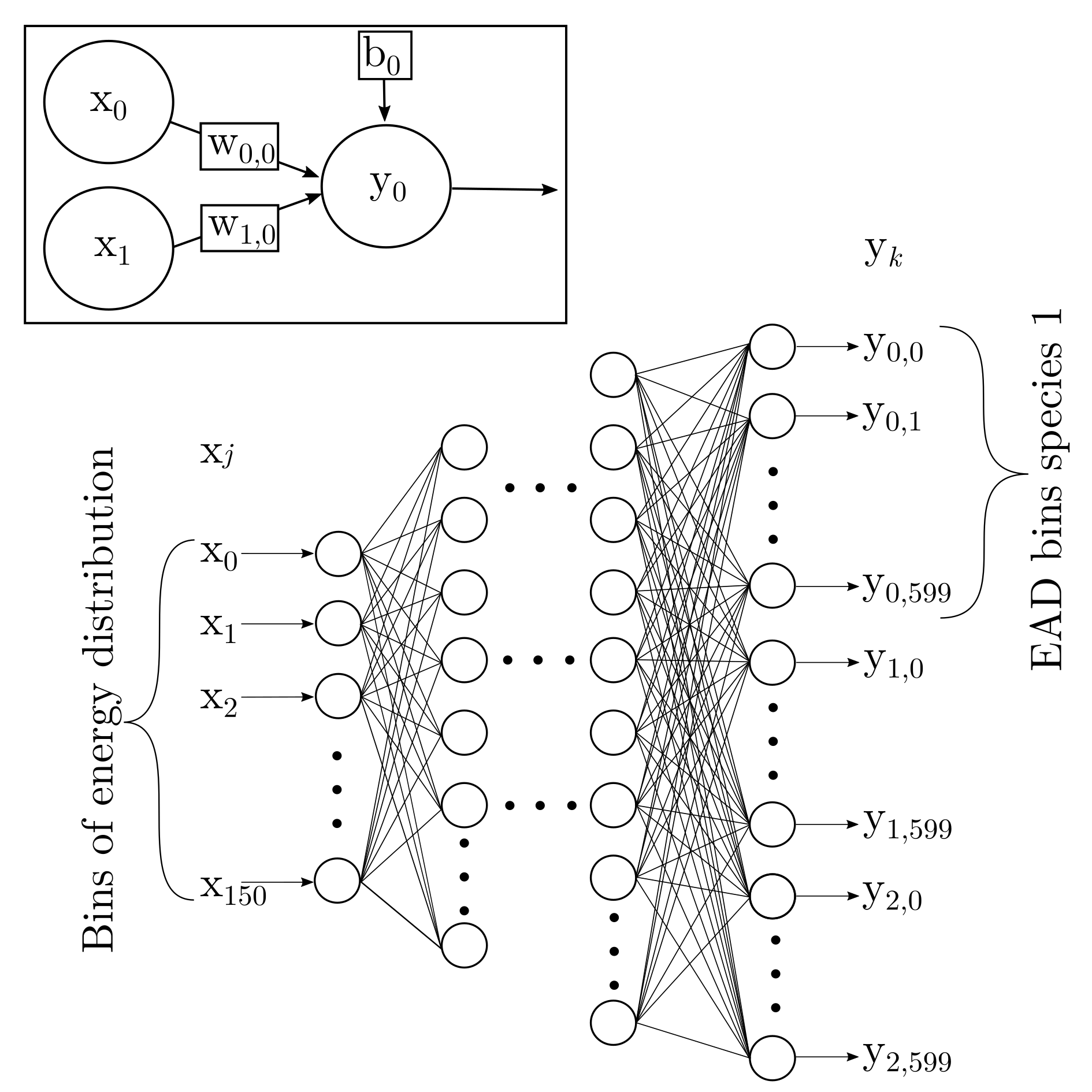}
\caption{Conceptual schematic of the utilized ANN.}
\label{fig:ANN}
\end{center}
\end{figure}

In MLP networks nodes are arranged in layers. In a fully connected network every node of a layer is connected to every node in the neighboring layers. The typical architecture of a feedforward neural network (like the one presented in this work) consists of a series of layers, each of which processing the (potentially nonlinear) outputs of the previous layer. The flow of information is directed from input to output and induces a causal relationship between them (cf.\ Figure~\ref{fig:ANN}). The layers between input and output layer are commonly referred to as hidden layers.\cite{pao_adaptive_1989,bishop_neural_1996} 

Linear hidden layers that may be introduced in addition to nonlinear activation layers increase the network complexity. However, linear algebra shows that adjacent linear layers can be combined to a single linear layer. Hence, nonlinear activation layers allow for the representation of dependencies, which are not linearly separable \cite{cybenko_approximation_1989}.

The number of nodes in the input and output layer of the network are defined by the specific problem to be approached. In the present scenario, they are specified by the bins of the respective histograms. The network input vector $x_j \, \hat{=} \, f_{\text i}[l]$ is given by the discrete energy distribution of the incident species with $\dim(x_j) = \dim(E_l) = 151$ input bins (which directly corresponds to the TRIDYN input). The output vector $y_k$ consists of two-dimensional EAD histograms $f_{\text{o},s}[m,n]$ of all considered species $s$ (i.e., reflected Ar and sputtered Al and Ti). Hence, the output vector corresponds to a flattened and segmented one-dimensional data structure. Specifically, the EAD of each species is initially flattened with respect to surface normal cosine $c_n$ and subsequently the collection of EADs is segmented with respect to species $s$. With $\dim(E_m) = 30$, $\dim(c_n) = 20$, and 3 species, this results in $\dim(y_k) = 1800$ output bins. While the input and output dimensions may, in principle, be chosen arbitrarily, the above values have been chosen as a compromise between data reduction and physical accuracy.

The conceptual structure of the resulting ANN with respect to the data structure is shown in Figure~\ref{fig:ANN}. The specific choice of network hyperparameters (e.g., number of layers, nodes per layer, activation functions) are discussed in Sections~\ref{sec:hyperparameters} and \ref{sec:results_hyperparameters}. Following the supervised learning concept, the network has to be trained on a previously known data set $N$ to make useful predictions. This set $N$ is composed of an input set $X$ containing reference inputs $x_j^{\,\prime} \in X$ and an output set $Y$ containing the corresponding reference outputs $y_k^{\,\prime}(x_j^{\,\prime}) \in Y$. The latter were \textit{a priori} calculated using TRIDYN.

For training the network, all weights and biases are randomly initialized. These are subsequently optimized using backpropagation by minimizing the loss function $e^2$. In this work, the mean square error $e^2(y_k, y_k^{\,\prime})$ of the predicted output $y_k(x_j^{\,\prime})$ and the ideal output $y_k^{\,\prime}(x_j^{\,\prime})$ was used. The average over all samples of a set defines the global error $\langle e^2(y_k, y_k^{\,\prime}) \rangle$. Minimization of this error is called training. One iteration of the training algorithm is called an epoch. Finally, the trained network can be used to make predictions $y_k(x_j)$ for previously unknown inputs $x_j \notin X $. \cite{bishop_neural_1996,pao_adaptive_1989}

To set up and train the ANN, the Tensorflow framework 1.8.0 and the Keras API 2.2.0 were used \cite{abadi_tensorflow:_2016-1,chollet_keras:_2015}. In addition, graphics processing unit acceleration was employed using CUDA 9.1 and two Tesla K40C GPUs \cite{nickolls_scalable_2008}.

\subsection{Generalization/Overfitting}
\label{sec:overfitting}

Generalization is an ANNs ability to handle novel data. It manifests in meaningful predictions (minimal loss), which often expose minimum statistical residual variation (i.e., noise). Minimal losses may be encountered in training of noisy data also due to overfitting of the weights and/or excessive model complexity. That is, a network's degrees of freedom may allow for an accurate representation of the training data -- including its statistical noise. Notably, this adaption of the network to faulty data often implies bad generalization \cite{bishop_neural_1996}. The small data set used in this work makes the training procedure especially prone to overfitting. As a remedy the data set $N$ consisting of 439 samples is partitioned into mutually exclusive training, validation, and test sets $N=\left \{ N_{\text{train}}, N_{\text{val}}, N_{\text{test}}\right \}$ with $N_{\text{train}} \cap N_{\text{val}} \cap N_{\text{test}} = \emptyset$. This very common approach allows for the evaluation of the network performance on statistically independent data sets, as follows:
\begin{itemize}
\itemsep-1ex
\item Training set $N_{\text{train}}$ (80\% of samples) is used to fit the model.
\item Validation set $N_{\text{val}}$ (10\% of samples) is used to provide an unbiased evaluation of the fit on the training set. This is used for tuning the model hyperparameters (cf.\ Sections~\ref{sec:hyperparameters} and \ref{sec:results_hyperparameters}) or as reference for early stopping (following paragraph).
\item Test set $N_{\text{test}}$ (10\% of samples) is used to provide an unbiased evaluation of the final model fit on the training set.
\end{itemize}

\begin{figure}[t]
\begin{center}
\includegraphics[width=8cm]{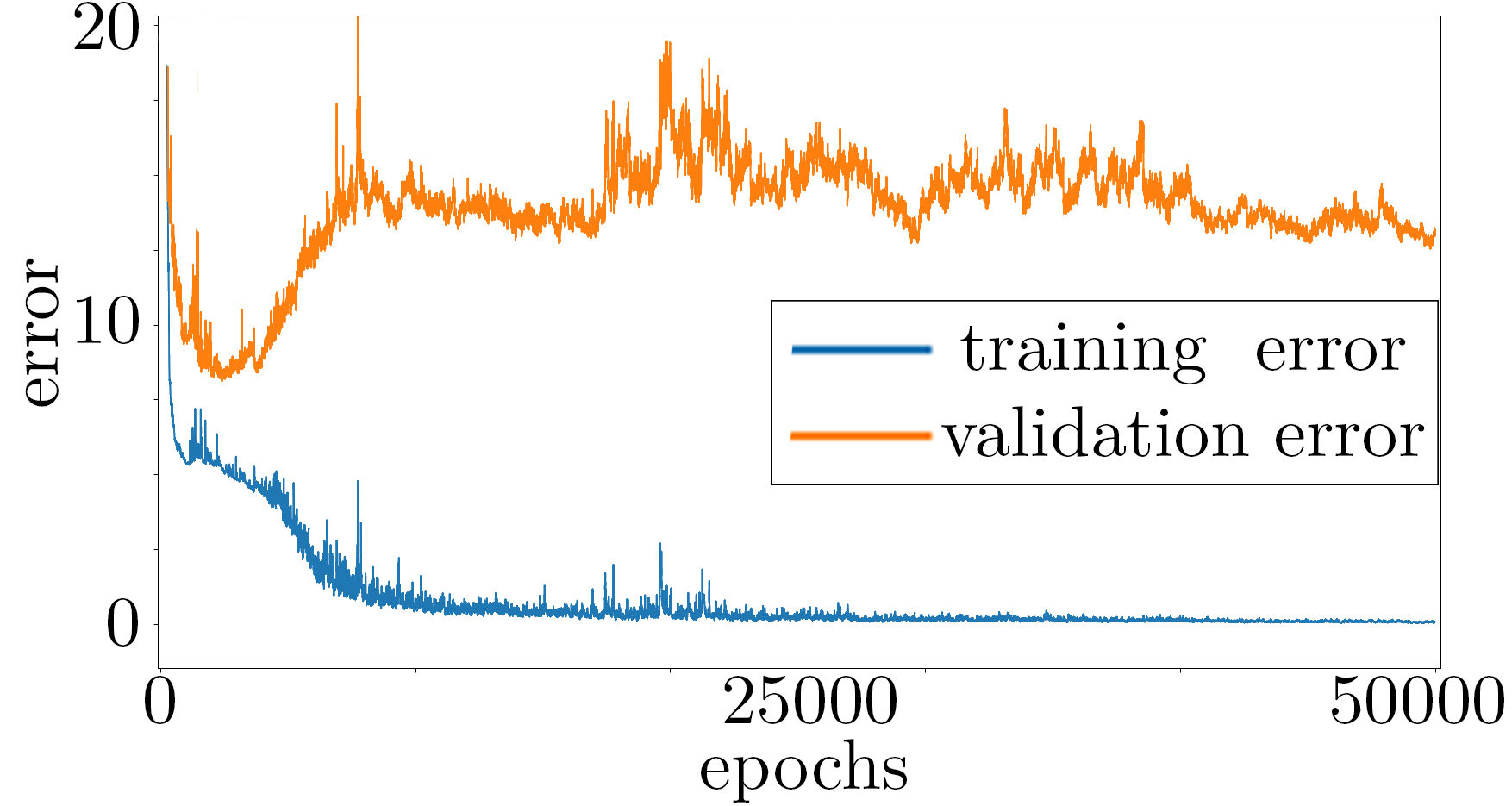}
\caption{Training and validation loss during the fitting procedure.}
\label{fig:noreg}
\end{center}
\end{figure}

Figure~\ref{fig:noreg} shows the evolution of training and validation losses over 50000 epochs for an exemplary training. Evidently, the training loss is continually declining, while the validation error exposes a local minimum after approximately 2600 epochs. This is a direct consequence of overfitting the data. To alleviate this effect, early stopping and regularization have been implemented. Precisely, in early stopping the training is stopped when the validation error stagnates or increases, while the training error keeps declining. In addition, regularization adds a penalty to the loss function, giving preference to well-generalizing network configurations. \cite{pao_adaptive_1989,bishop_neural_1996} The specific parameters are detailed in Section~\ref{sec:results_hyperparameters}.

\subsection{Network structure and hyperparameters}
\label{sec:hyperparameters}

The predictive capabilities of an ANN are significantly influenced by its structure and training configuration. The specifying parameters are commonly referred to as hyperparameters. A non-comprising list of the most relevant hyperparameters, their variation in the subsequent study, and the finally used values are listed in Table~\ref{table:finalANN}. While not all of the ones listed are introduced or detailed in this work, comprehensive information can be found in \cite{pao_adaptive_1989,bishop_neural_1996,reed_neural_1999}.

\begin{table}[t!]
\caption{ANN hyperparameters (subdivided into topology and learning).}
\label{table:finalANN}
\begin{center}
\begin{tabular}{ l c  c  c }
\hline
hyperparameter & variation & ~ suboptimal ~ & final \\
\hline
no. of input nodes & - & 151 & 151 \\
no. of output nodes & - & 1800 & 1800 \\
no. of hidden layers & [1, 3, 5, 7] & 5 & 3 \\
no. of nodes per layer & [500, 1000, 2000] & 1000 & 1000 \\
activation function & [sigmoid, tanh] & tanh & sigmoid \\
no. of activation layers \space\space & [1, 3, 5]$^\ast$ & 5 & 3 \\
\hline \hline
learning rate & - & 0.1 & 0.1 \\
decay & - & 0.1 & 0.1 \\
momentum & - & 0.1 & 0.1 \\
patience & [0, 500, 1000] & 1000 & 500 \\
regularization type & [L1, L2] & L1 & L1 \\
regularization factor & [0, 0.1, 0.5, 1] & 0 & 0.1\\
\hline
\multicolumn{4}{p{10cm}}{\linespread{.6}\small{$^\ast$ The number of activation layers is always chosen to be maximal, as far as the total number of hidden layers allows.}\linespread{1}}

\end{tabular}
\end{center}
\end{table}

Since the input and output vectors only define the outer layers of the network, the remaining network structure and most hyperparameters are not predefined by the data structure. Consequently, an optimal set of hyperparameter has to be found by conducting a parameter study. K-fold cross validation has been used to determine the validation error of the different combinations of the hyperparameters \cite{pao_adaptive_1989,bishop_neural_1996,zwillinger_standard_2002} and to determine the optimal set.

The design requirements are that the precision of predictions should be significantly improved compared to simplified methods (e.g., fitted analytical expressions) mentioned above. Since the trained model is intended for run-time evaluation, the prediction time must be considered as well. The computation time of kinetic simulations varies significantly, ranging from hours to weeks. Therefore, the additional prediction time should be evaluated relative to the total evaluation time per time step. For three-dimensional Monte Carlo simulations of a research scale PVD process evaluated on a parallel cluster with 80 CPU cores \cite{trieschmann_transport_2015}, the total evaluation time per time step is approximately $\Delta t_\text{eval} \approx 5$~s. In this case, approximately $N_\text{pred} \approx 500$ particles impinging the target surface have to be evaluated per time step. Hence, the total prediction time per time step should be $\Delta t_\text{pred} \ll \Delta t_\text{eval} / N_\text{pred} \approx 10$~ms. The prediction performance is characterized and shown to fulfill this requirement in a detailed hyperparameter analysis presented in following Section~\ref{sec:results_hyperparameters}.

\section{Results}
\label{sec:results}

As apparent from the previous discussion, a large variety of network configurations is possible even for the conceptually simple class of MLP networks. Not only the network structure, but also the specific implementation of the nodes (e.g., activation function) and choice of global learning parameters (e.g., decay) offer flexibility in network design. A careful selection of corresponding hyperparameters is required for optimal prediction.

\subsection{Hyperparameter space and validation}
\label{sec:results_hyperparameters}

\begin{figure}[t]
\begin{center}

\begin{subfigure}{8cm}
\begin{center}

\includegraphics[width=8cm]{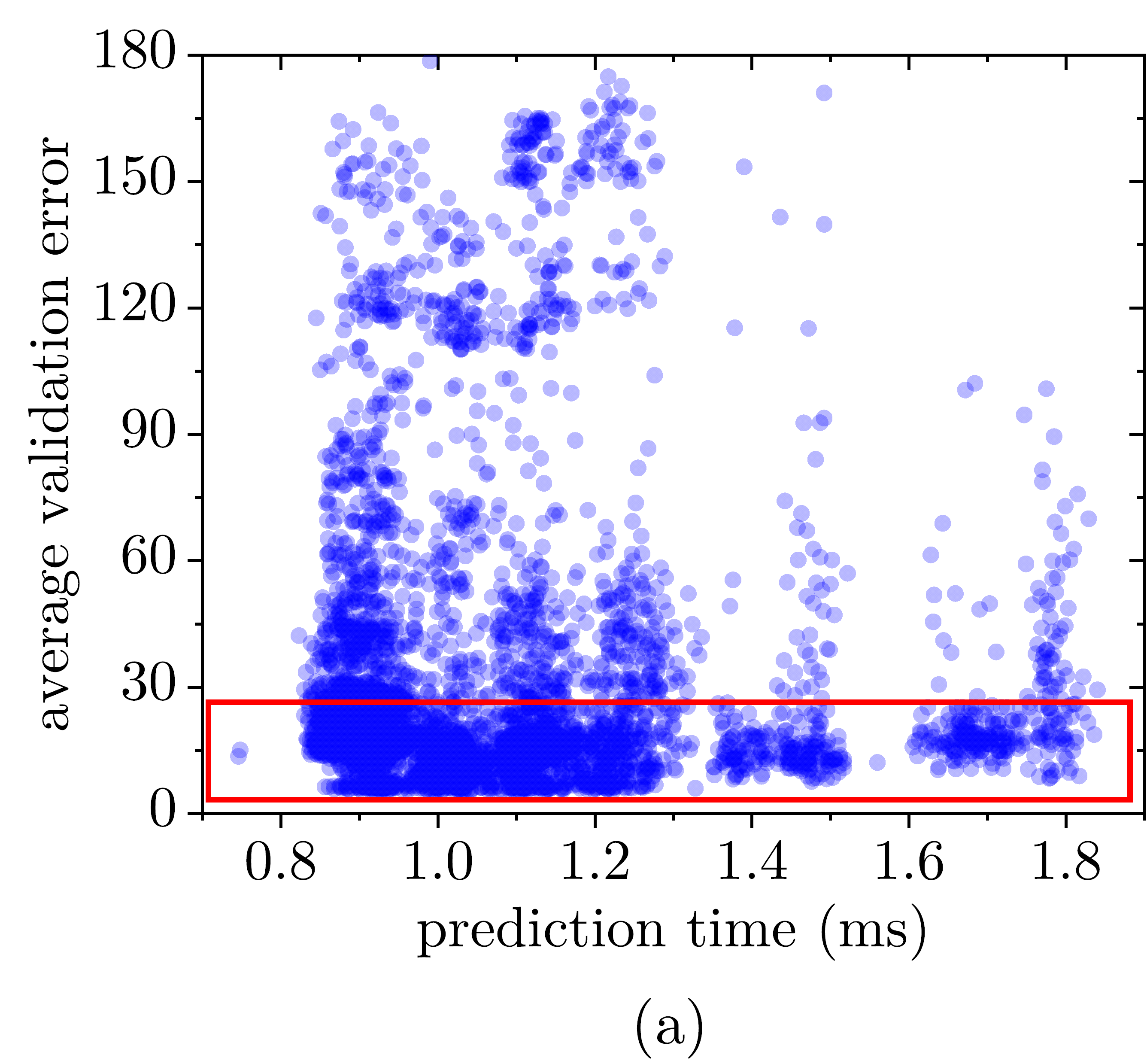}
\caption{\vspace{-4cm}}
\label{fig:total}
\end{center}
\end{subfigure}
\begin{subfigure}{8cm}
\begin{center}
\includegraphics[width=8cm]{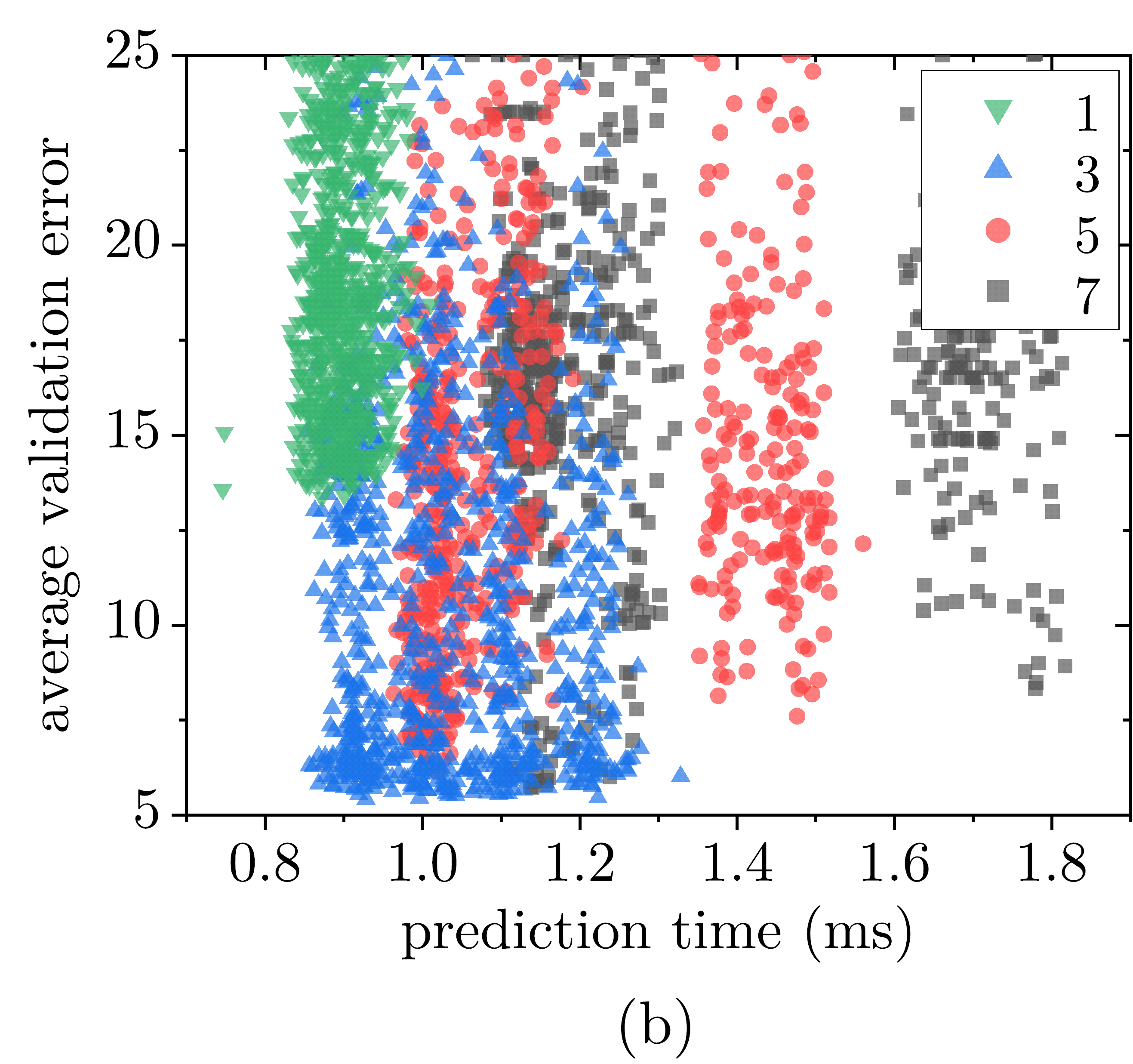}
\caption{\vspace{-4cm}}
\label{fig:varlayers}
\end{center}
\end{subfigure}

\begin{subfigure}{8cm}
\begin{center}
\includegraphics[width=8cm]{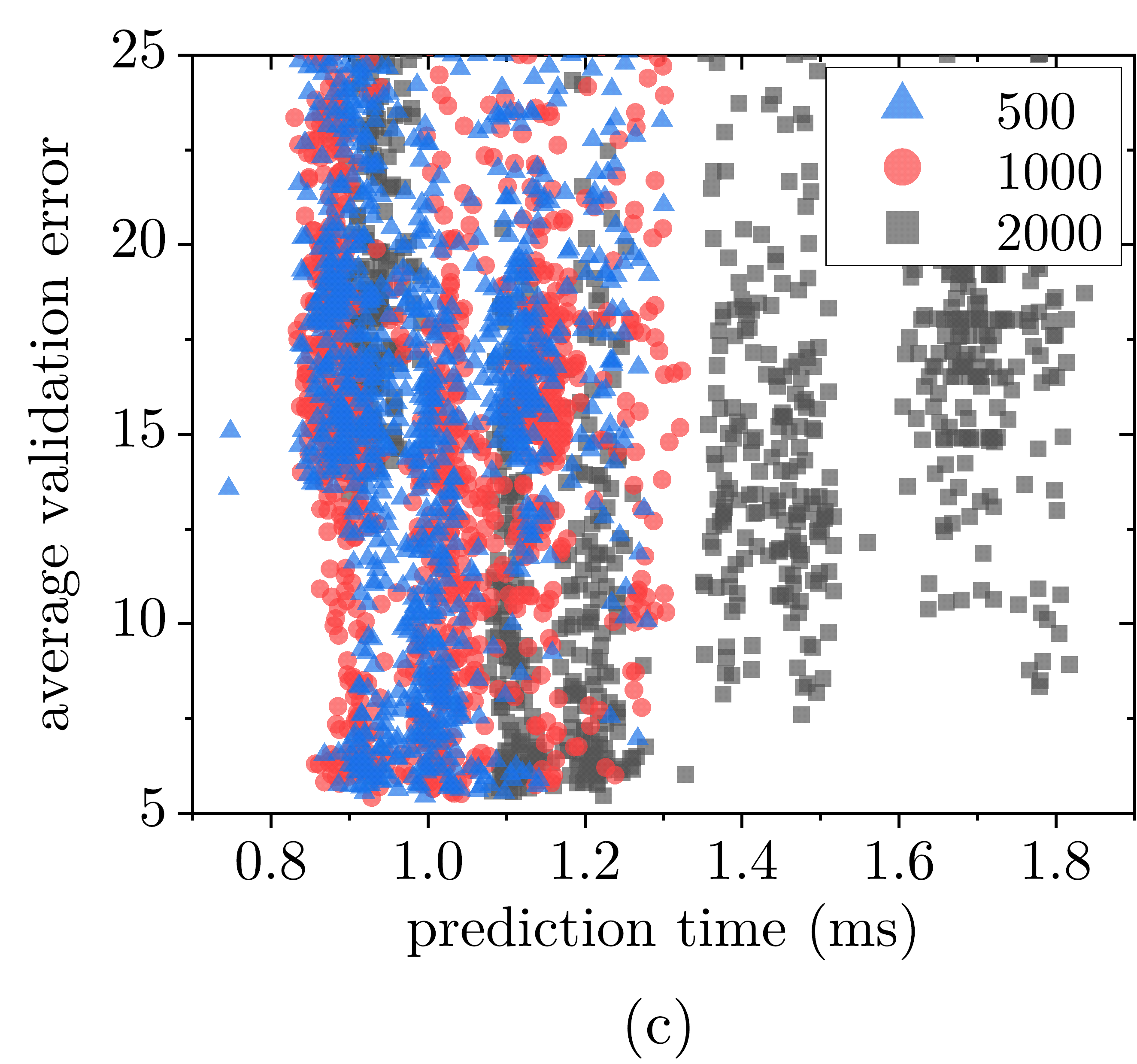}
\caption{\vspace{-4cm}}
\label{fig:varnodes}
\end{center}
\end{subfigure}
\begin{subfigure}{8cm}
\begin{center}
\includegraphics[width=8cm]{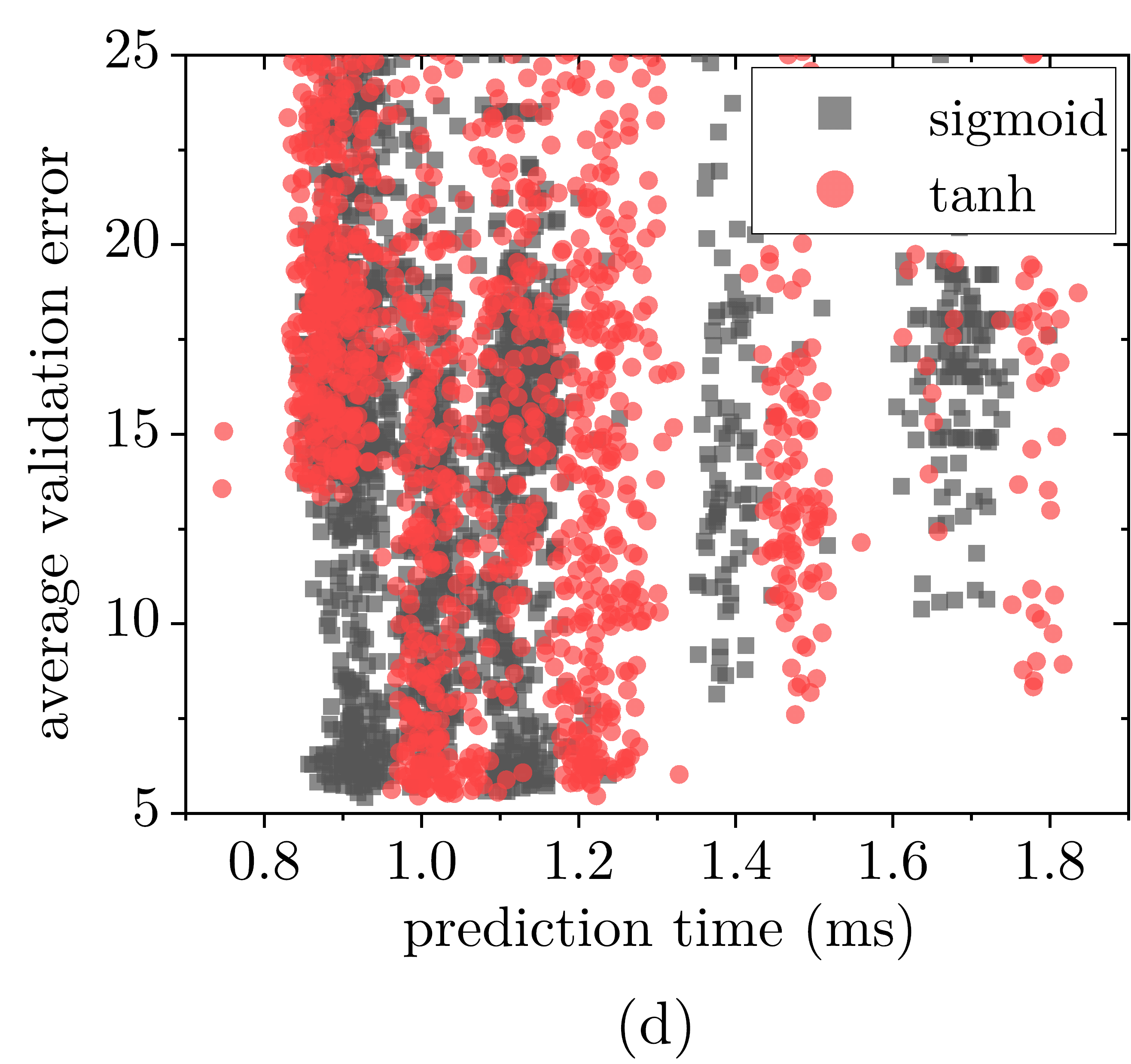}
\caption{\vspace{-4cm}}
\label{fig:varactivation}
\end{center}
\end{subfigure}

\end{center}
\vspace{-20pt}
\caption{(a) Performance of converged hyperparameter sets, (b) influence of number of hidden layers on performance for subset, (c) influence of number of nodes per layer on performance for subset, (d) influence of activation function on performance for subset.}
\end{figure}

It is instructive to rigorously map out the hyperparameter space before considering the optimal case used for final evaluation. This was done with a variation of parameters as listed in Table~\ref{table:finalANN}. It was limited to the presented value ranges due to the amount of combinations to be trained (576 in total) and the corresponding computational effort. The average validation error $\langle e_{\text{val}}^2 \rangle$ versus the prediction time $\Delta t_\text{pred}$ of all converged hyperparameter sets is depicted in Figure~\ref{fig:total}. Because predictions with low validation error are generally preferable, the correspondingly marked section (red box) is further investigated. The distinction which defines a ``low error'' is arbitrary and only for illustration purposes. Yet, the substantial disparity from $\langle e_{\text{val}}^2 \rangle \approx 5$ to 180 underlines the importance of a proper hyperparameter choice.

Figure~\ref{fig:varlayers} shows the performance of trained ANNs segregated by the number of hidden layers. The networks containing only 1 hidden layer (green) clearly tend to perform faster than those containing more layers. This is expected, as the number of floating-point operations increases with the number of hidden layers (for a constant number of nodes per layer). Additionally, the causal relationship between the layers requires sequential computation, thereby limiting acceleration through parallelized computing. However, it is also observed that the minimum validation error of these networks is significantly larger than for networks with more hidden layers. This is due to the circumstance that the model complexity is insufficient to adequately model the higher-order input-output relation \cite{haykin_neural_2008}. In contrast, those networks with 3 or more hidden layers (and appropriate regularization) show comparably small validation errors, while their prediction time analogously increases with the number of hidden layers.

A notable aspect relates to the number of nonlinear activation and hidden layers. Networks with 5 and 7 hidden layers both have a maximum number of 5 nonlinear activation layers, since a further increase compromises convergence during training. These networks differ in the number of linear layers, which for the present study manifests in two aspects: (i) an unchanged validation error and (ii) an extended prediction time. This observation is reasoned by the increased but linear network complexity as discussed in Section~\ref{sec:ANN}.

When comparing the number of nodes per layer as shown in Figure~\ref{fig:varnodes}, no significant trend in the minimal reachable validation error is apparent (i.e., the model complexity is appropriate for all considered cases). Only a vague trend is noticeable concerning the prediction time: It may be inferred to increase from 500 (blue) over 1000 (red) to 2000 (black) nodes per layer, but a significant overlap renders a general statement difficult. The increased prediction time is most evident for 2000 nodes per layer, reaching prediction times of up to $\Delta t_\text{pred} \approx 1.8$~ms. This is again reasoned by the increase in floating-point operations required with increasing number of nodes per layer (for a constant number of hidden layers).

As evident from Figure~\ref{fig:varactivation}, the activation function has an important influence on the network performance as well. In fact, the variability of the sigmoid function, with its property to be non-negative, proves most computationally efficient for a comparable structural network complexity (i.e., same number of hidden layers and number of nodes per layer). This trend is clearly observed for example in the case of 2000 nodes per layer and 5 hidden layers. The column at about 1.4 to 1.5 ms shows a clear segregation between tanh and sigmoid based networks, with the latter resulting in faster prediction times. However, while this can be concluded for the specific case investigated in this work, general statements for other problem types are difficult \cite{haykin_neural_2008}.

An additional aspect related to a network's ability to generalize after training is the strong link of the achievable validation error to the applied regularization. Specifically, the quality of generalization, and therewith the ability to make predictions on unknown input data, was found to be influenced substantially. Without regularization the trained ANNs were prone to overfitting, while regularization during training effectively suppressed false adaption to noise for the same network. Regularization was, therefore, determined to be crucial, due to the comparably small data set $N$ and the intentionally entailed statistical noise. However, from the obtained data (not shown) no general trend for the applied regularization type and penalty factor could be inferred. This suggests that the existence of regularization, rather than its type or penalty factor, determine a successful adaptation.

The goal to utilize the trained ANN in the frame of a Monte Carlo gas-phase simulation imposes the previously mentioned requirement on prediction time. It is evident from Figure~\ref{fig:total} that essentially all considered networks fulfill this requirement with $\Delta t_\text{pred} \approx 0.8~\text{to}~1.8~\text{ms} \ll 10$~ms. Nevertheless, a minimal prediction time is preferable given comparable validation errors and is correspondingly chosen. Combining the presented insights suggests the set of optimal hyperparameters: 3 hidden layers for adequate model complexity and 1000 nodes per layer as well as sigmoid activation function for optimal prediction time. Final parameters are listed in Table~\ref{table:finalANN}. The optimized network yields an average validation error $\langle e_{\text{val}}^2 \rangle = 5.83$ and a prediction time of $\Delta t_\text{pred} = 0.87$~ms.

\subsection{Demonstration and test}
\label{sec:results_demonstration}

The previously optimized ANN was used to generate predictions for Ar projectiles hitting a Ti-Al target surface using a randomly selected, previously unknown input energy distribution of the test set $N_\text{test}$. In the following example, a Gaussian energy distribution peaked with $\sigma^2 =0.3$~eV at $E = 590$~eV is further investigated.

\begin{figure}[b!]
\includegraphics[width=16cm]{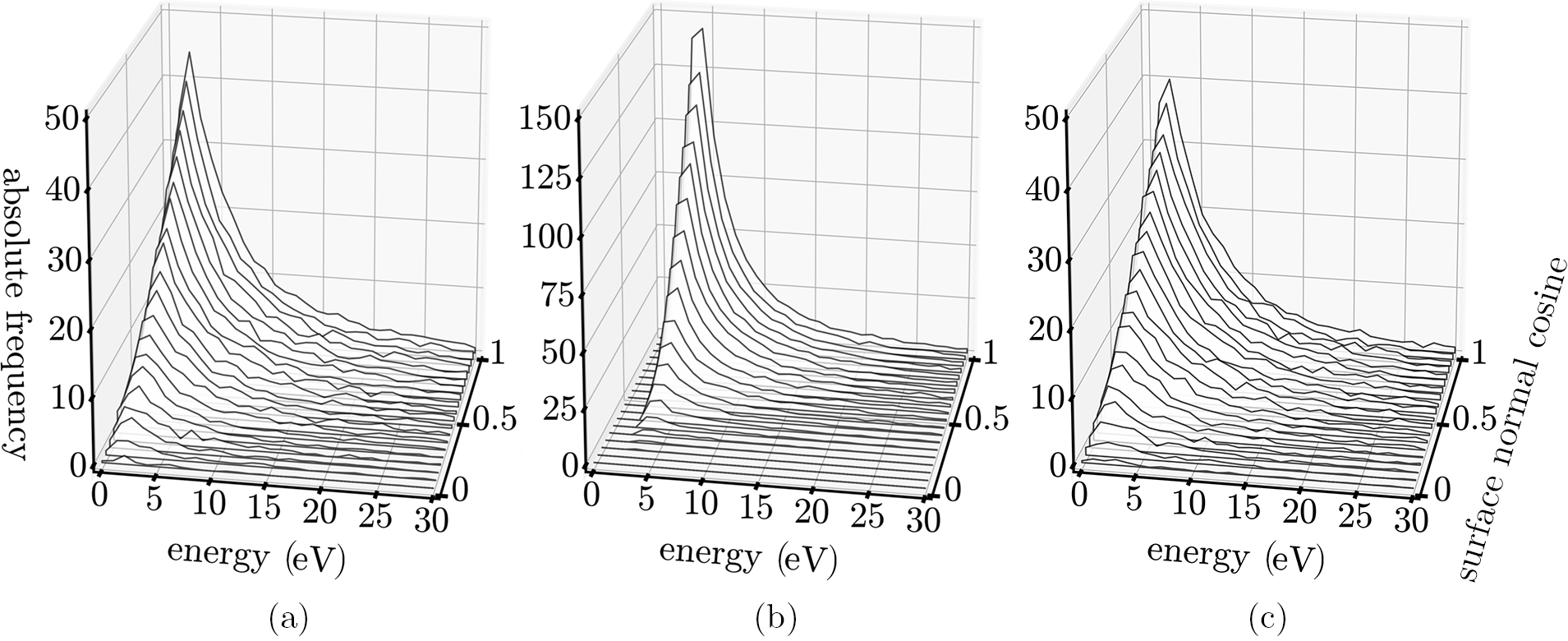} \\
\includegraphics[width=16cm]{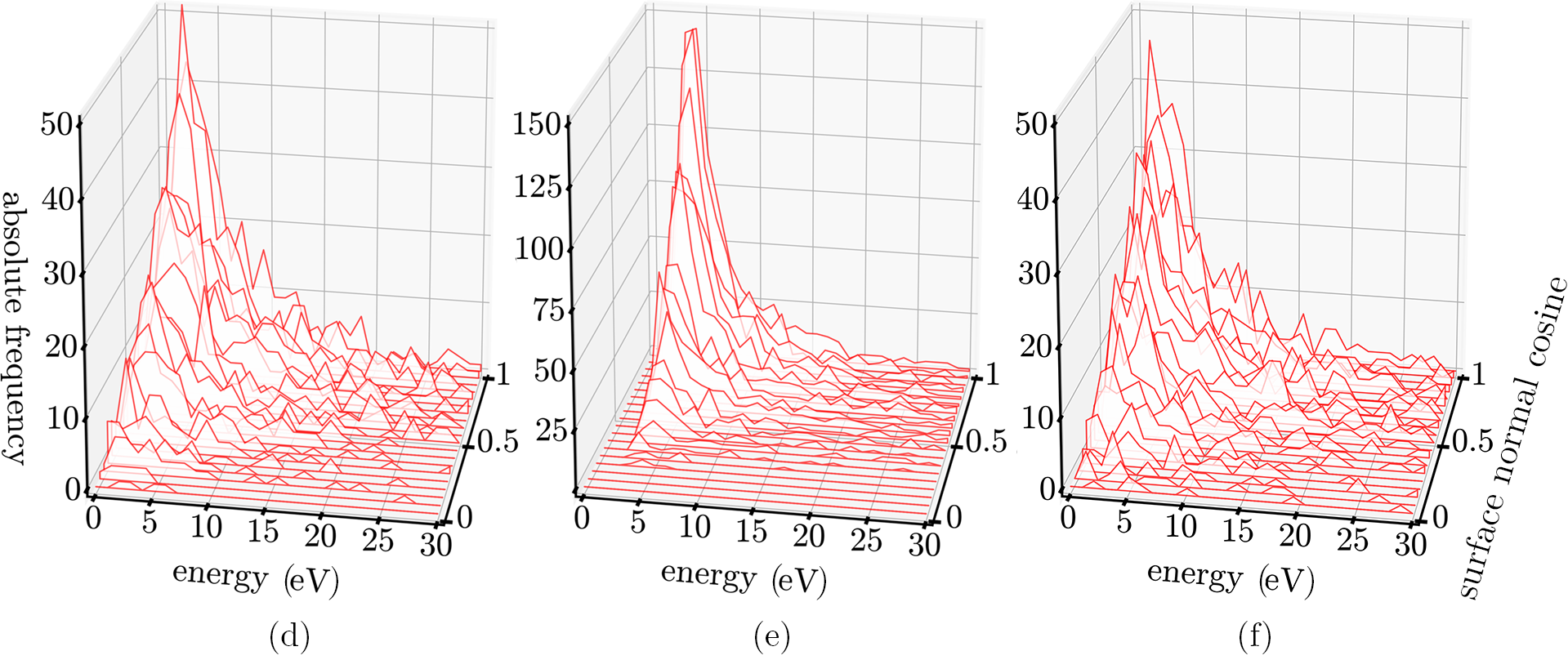}
\caption{Test case comparison (a) Al EAD prediction, (b) Ar EAD prediction, (c) Ti EAD prediction, (d) Al EAD reference, (e) Ar EAD reference and (f) Ti EAD reference.}
\label{fig:bench1_EAD} 
\end{figure}

Figure~\ref{fig:bench1_EAD} shows the correspondingly predicted EADs of Ar, Al, and Ti together with reference TRIDYN results as used for training (low statistical quality). The predictions (top, black) show qualitative and quantitative agreement with the reference solution (bottom, red). This is confirmed by the respective mean value $\langle y_k \rangle$ of the output vectors and their variance $\sigma^2$ , given by

\begin{equation}
\label{eq:variance}
\sigma^2(y_k) = \frac{1}{\dim(y_k)} \sum_{k=0}^{\dim(y_k)-1} (y_k-\langle y_k \rangle)^2.
\end{equation}

In the presented case these value are: $\langle y_k \rangle = 5.71 \approx \langle y_k^{\,\prime} \rangle = 5.67$ and $\sigma^2(y_k) = 150.51 \approx \sigma^2(y_k^{\,\prime}) = 158.90$. However, while the distributions' shapes and magnitudes are accurately reproduced (EADs are not normalized), a higher statistical quality of the predicted data as compared to the reference solution is observed. This is an inherent consequence of the ANN's ability to generalize, whereby random and false noise contributions are effectively ignored.

\begin{figure}[t]
\includegraphics[width=16cm]{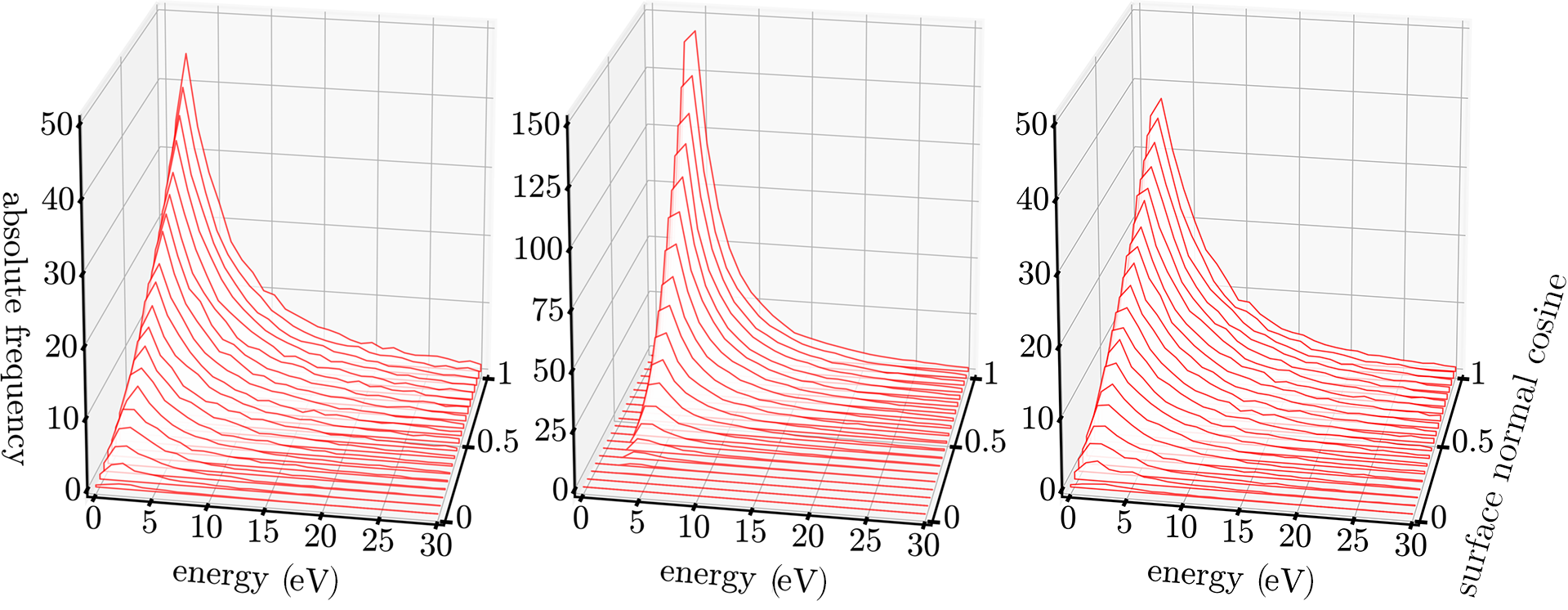}
\caption{EADs of the high statistical quality TRIDYN reference evaluated for $N_\text{sp} = 10^6$ projectiles. Al (left), Ar (center), and Ti (right).}
\label{fig:high_low_statistics}
\end{figure}

To quantify the prediction quality more accurately, a reference TRIDYN simulation with high statistical quality ($N_\text{sp} = 10^6$ instead of $10^4$ incident projectiles) was performed. This high quality data set $y_k^{\,\prime\prime}$ provides a low-noise reference for both the prediction $y_k$ and the training reference $y_k^{\,\prime}$. It is depicted in Figure~\ref{fig:high_low_statistics}, which shows the corresponding EADs for Al (left), Ar (center), and Ti (right). The coefficient of determination $R^2$ gives a statistical measure of the quality of a regression model. Using $y_k^{\,\prime\prime}$ as reference, it is defined as \cite{draper_applied_1998}
\begin{equation}
\label{eq:coefficient_of_determination}
R^2=1-\frac{SS_\text{res}}{SS_\text{tot}}
\end{equation}
with the residual sum of squares
\begin{equation}
\label{eq: The_sum_of_squares_of_residuals}
SS_\text{res} = \sum_{k=0}^{\dim(y_k)-1} \left( y_k^{\,\prime\prime} - y_k \right)^2 =\dim(y_k) \,  e^2(y_k, y_k^{\,\prime\prime})
\end{equation}
and the total sum of squares of $y_k^{\,\prime\prime}$
\begin{equation}
\label{eq:total_sum_of_squares}
SS_\text{tot} = \sum_{k=0}^{\dim(y_k^{\,\prime\prime})-1} \left( y_k^{\,\prime\prime} - \langle y_k^{\,\prime\prime} \rangle \right)^2.
\end{equation}
Notably, only for $y_k^{\,\prime}$ as reference $SS_\text{res} \, \hat{=} \, \dim(y_k^{\,\prime}) \, e^2(y_k, y_k^{\,\prime})$ during training/validation and $SS_\text{tot} \, \hat{=} \, \dim(y_k^{\,\prime}) \, \sigma^2(y_k^{\,\prime})$ is proportional to the previously evaluated variance. Also note that the calculated values comprise the contributions of all EADs (flattened output vector $y_k$).

In the investigated case, equations \eqref{eq:coefficient_of_determination}--\eqref{eq:total_sum_of_squares} yield $R_\text{pred}^2=99.85\%$. It stands out that although the network was exclusively trained on high-noise data it shows excellent agreement with the high quality reference data. Interestingly, it even surpasses the limited accuracy case with $R_\text{ref}^2=96.12\%$, which proves that the effects of statistical noise have been mitigated significantly through generalization.

\begin{figure*}[b]
\centering
\includegraphics[width=.99\linewidth]{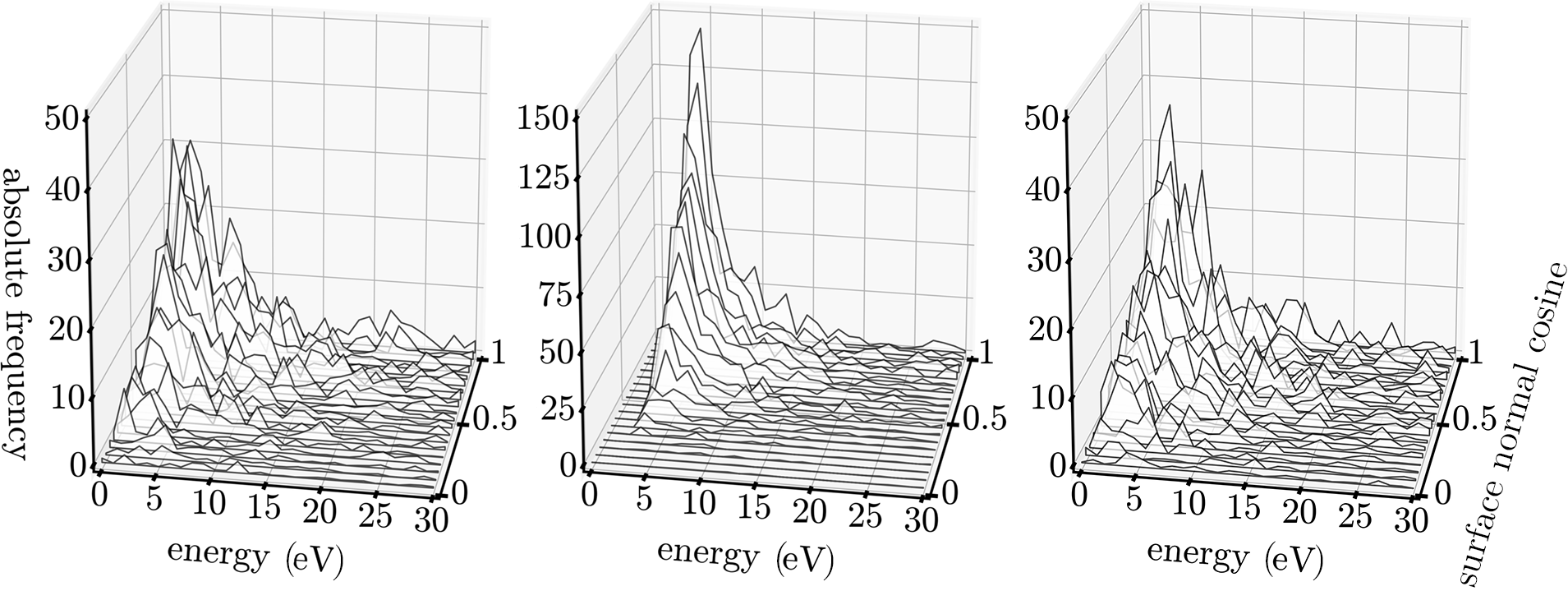}
\caption{Predicted EADs for suboptimal hyperparameters for Al (left), Ar (center), and Ti (right).} 
\label{fig:compare}
\end{figure*}

The importance of well-chosen hyperparameters that yield an ANN with good generalization capabilities becomes evident when comparing the prediction results of the optimal and a suboptimal set as presented in Figure~\ref{fig:compare}. This configuration was chosen arbitrarily from the remaining hyperparameter sets depicted in Figure~\ref{fig:total}. Its parameters are also included in Table~\ref{table:finalANN}. The suboptimal configuration yields a average validation error of 15.5 (approximately threefold increase) and produces predictions with significantly higher statistical fluctuations. This is verified by the coefficient of determination with respect to $y_k^{\,\prime\prime}$, which yields $R^2_\text{sub}=77.43\%$, significantly worse than the optimal set. From the network structure (cf.\ Table~\ref{table:finalANN}) it can be inferred that this is clearly due to failed generalization, rather than too simple network complexity.

\captionsetup[subfigure]{labelformat=parens}
\captionsetup[subfigure]{margin=3.0cm}
\captionsetup[subfigure]{skip=-0.1cm}
\subsection{Integrated distributions and sputter yield}
\label{sec:results_yield}

\begin{figure*}[b!]
\begin{center}
\begin{subfigure}{5.3333cm}
  \includegraphics[width=5.3333cm]{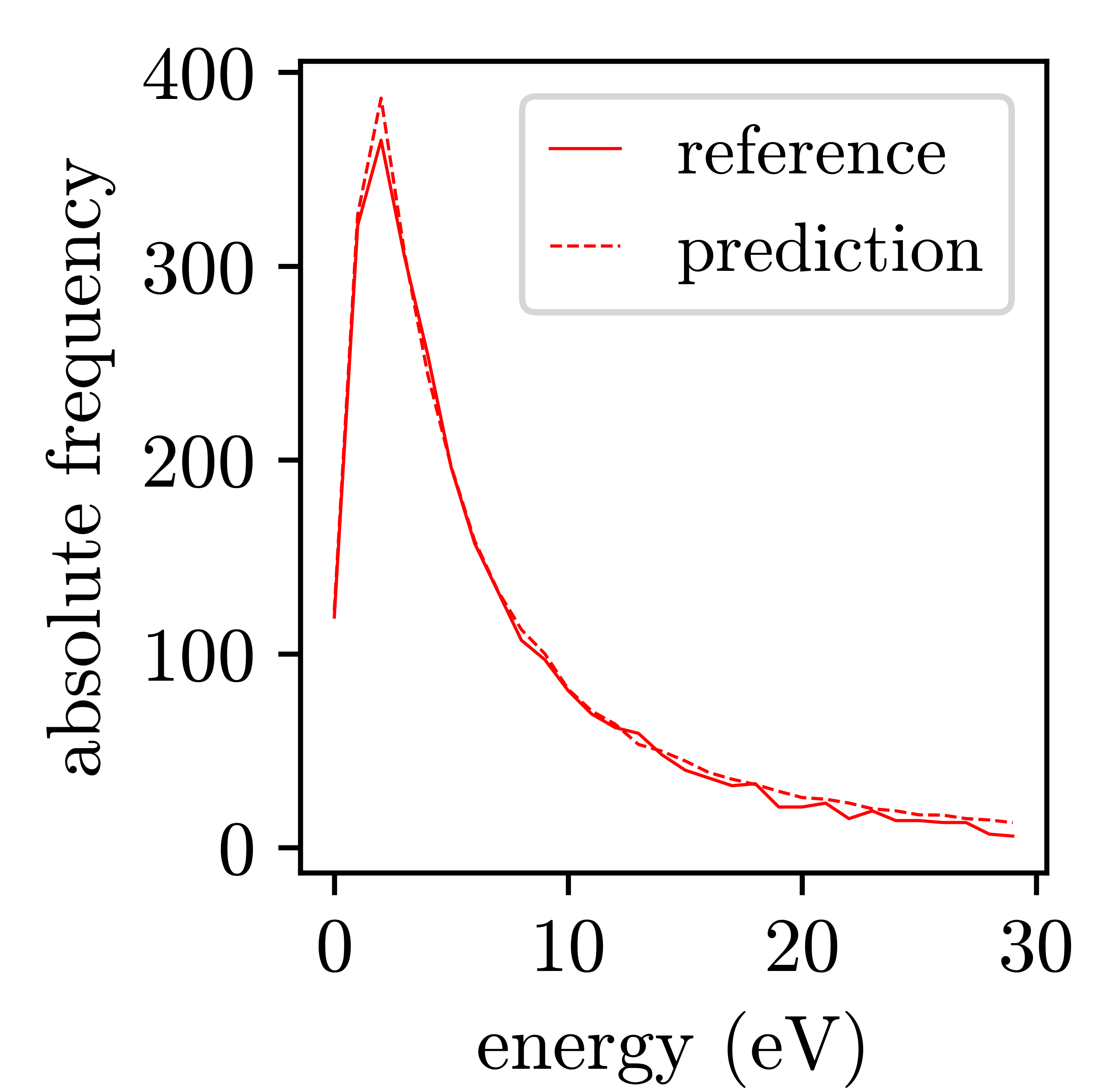}
  \caption{}
  \label{fig:bench1a}
\end{subfigure}%
\begin{subfigure}{5.3333cm}
  \includegraphics[width=5.3333cm]{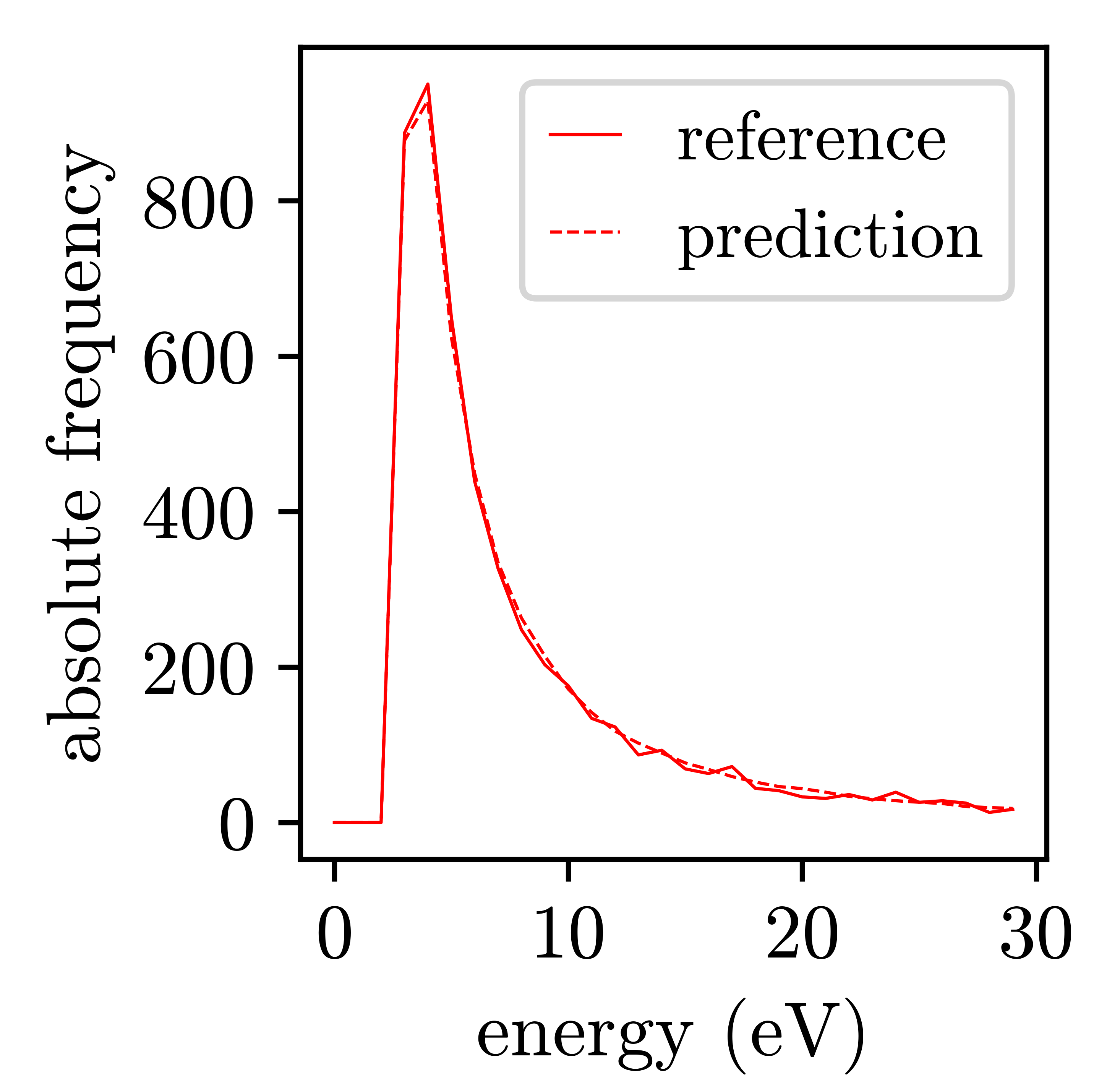}
  \caption{}
  \label{fig:bench1b}
\end{subfigure}%
\begin{subfigure}{5.3333cm}
  \includegraphics[width=5.3333cm]{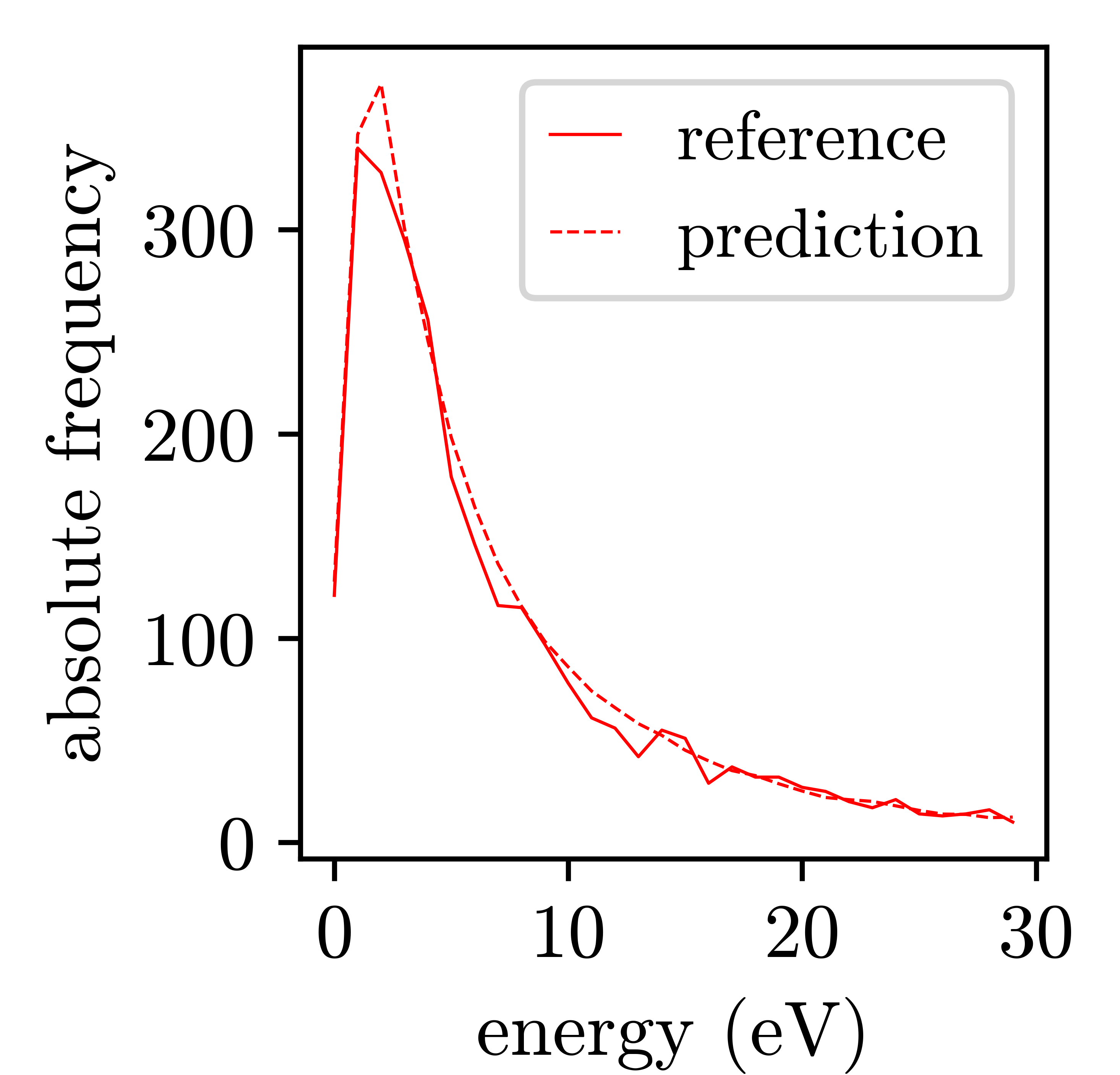}
  \caption{}
  \label{fig:bench1c}
\end{subfigure}
\begin{subfigure}{5.3333cm}
  \includegraphics[width=5.3333cm]{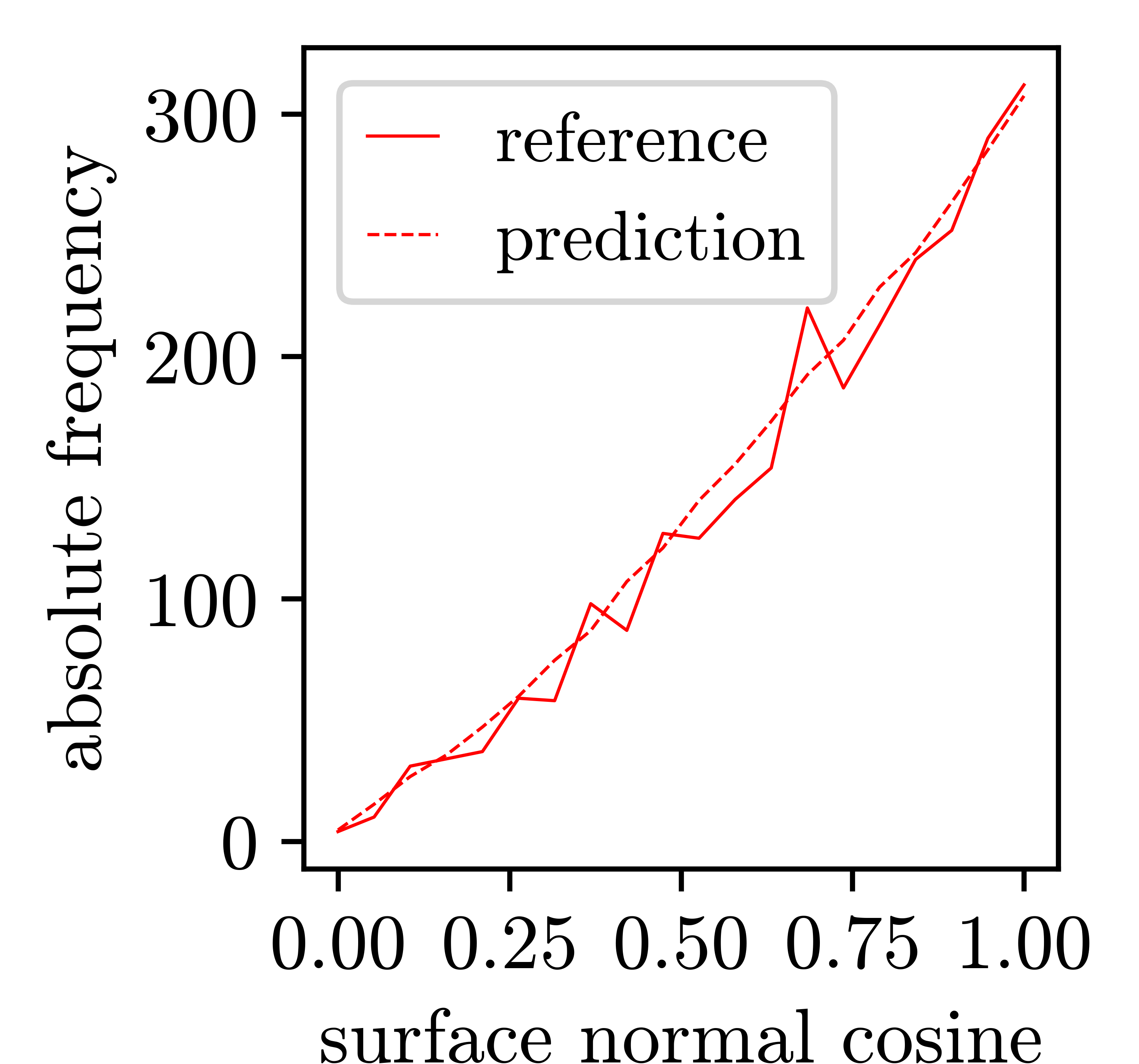}
  \caption{}
  \label{fig:bench1d}
\end{subfigure}%
\begin{subfigure}{5.3333cm}
  \includegraphics[width=5.3333cm]{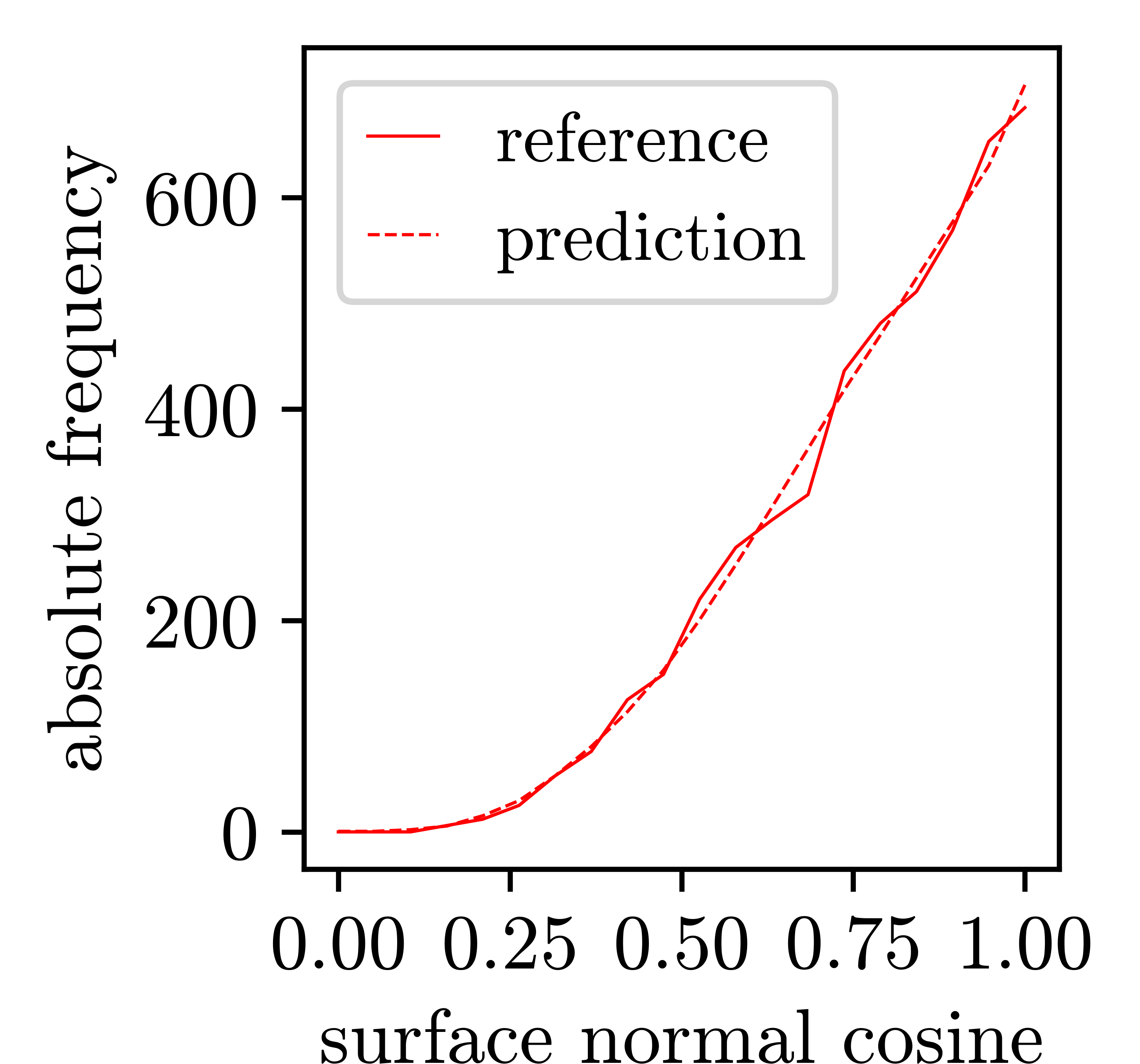}
  \caption{}
  \label{fig:bench1e}
\end{subfigure}%
\begin{subfigure}{5.3333cm}
  \includegraphics[width=5.3333cm]{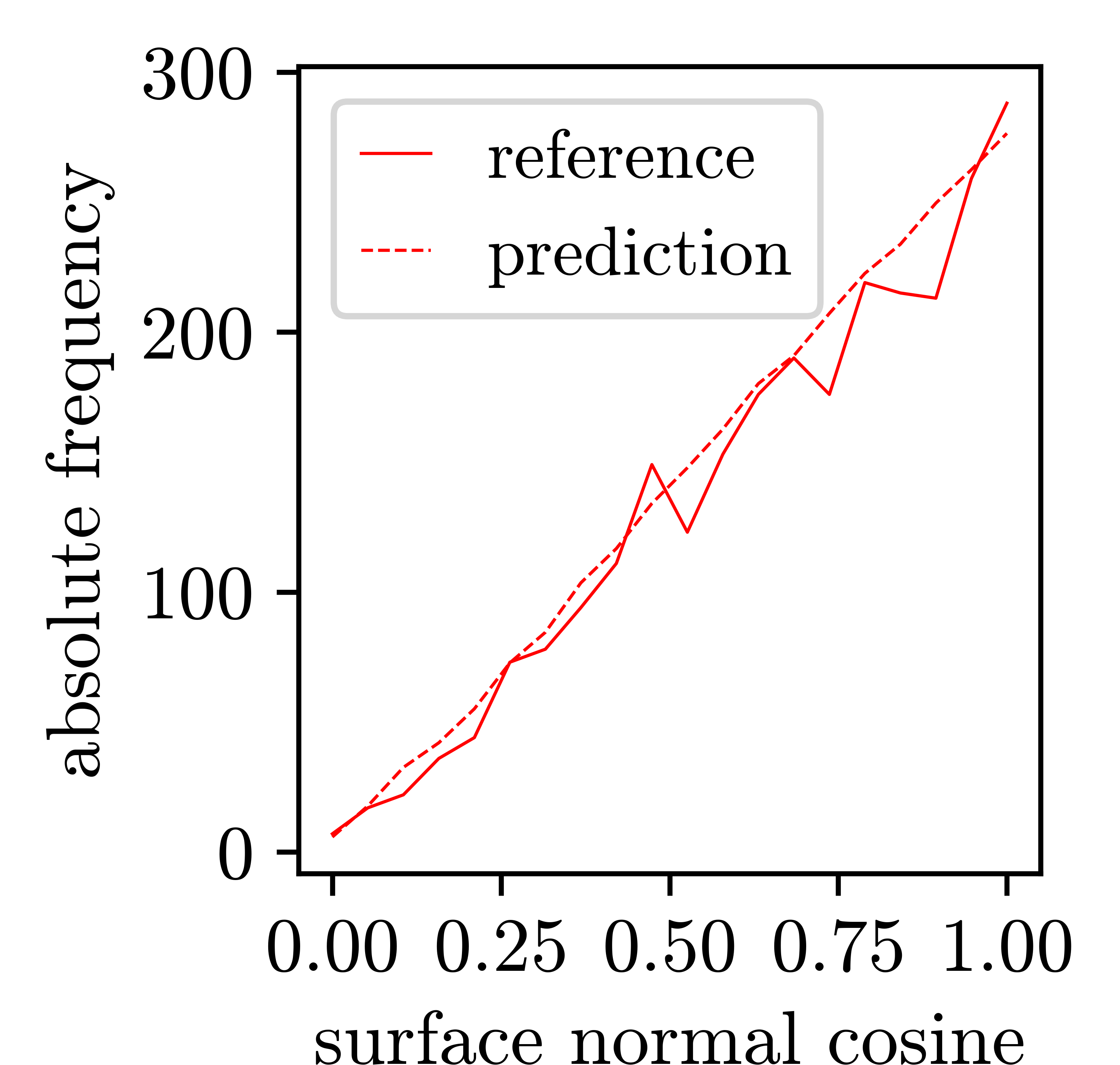}
  \caption{}
  \label{fig:bench1f}
\end{subfigure}

\caption{Predicted and reference (a) Al energy distribution, (b) Ar energy distribution,  (c) Ti energy distribution, (d) Al angular distribution, (e) Ar angular distribution and (f) Ti angular distribution.}
\label{fig:bench1}
\end{center}
\vspace{-20pt}
\end{figure*}

The validity of the predicted results can be further checked by considering the dependencies of integrated quantities (i.e., moments of the distribution) and comparing them against reference results. Integrating the EADs over polar angle or energy separates the energy and angular dependence of the distributions, respectively. The resulting energy distributions are shown in Figures~\ref{fig:bench1a} to \ref{fig:bench1c}, while the resulting angular distributions are depicted in Figures~\ref{fig:bench1d} to \ref{fig:bench1f}. The integration reduces the statistical scatter with increasing number of particles per bin. The smoothness of the prediction and the reference improve accordingly. For the energy distributions, the relative deviation between prediction and reference is 1.1\% on average and 15\% at maximum. For the angular distributions it is 3\% and 18.4\%, respectively. It should be noted that large relative deviations occur exclusively at the low magnitude outskirts of the distributions, which leads to the respective absolute deviation being negligibly small. Furthermore, analogous to the methodology in Section~\ref{sec:results_demonstration}, the respective coefficients of determination are $R^2_\text{pred, E}=99.95\%$ and $R^2_\text{ref, E}=99.47\%$ as well as $R^2_\text{pred, A}=99.96\%$ and $R^2_\text{ref, A}=99.28\%$ for energy and angular distributions. This further confirms that the previously observed limited agreement between $y_k^{\,\prime}$ and $y_k^{\,\prime\prime}$ is due to statistical perturbations, once more indicating the appropriateness of the optimal ANN's prediction.

\begin{figure}[b!]
\centering
\caption{Predicted and reference sputter yield (Al/Ti) and reflection coefficient (Ar) as a function of incident projectile energy using the mono-energetic distribution \eqref{eq:distribution1}.}
\label{fig:yield}
\includegraphics[width=8cm]{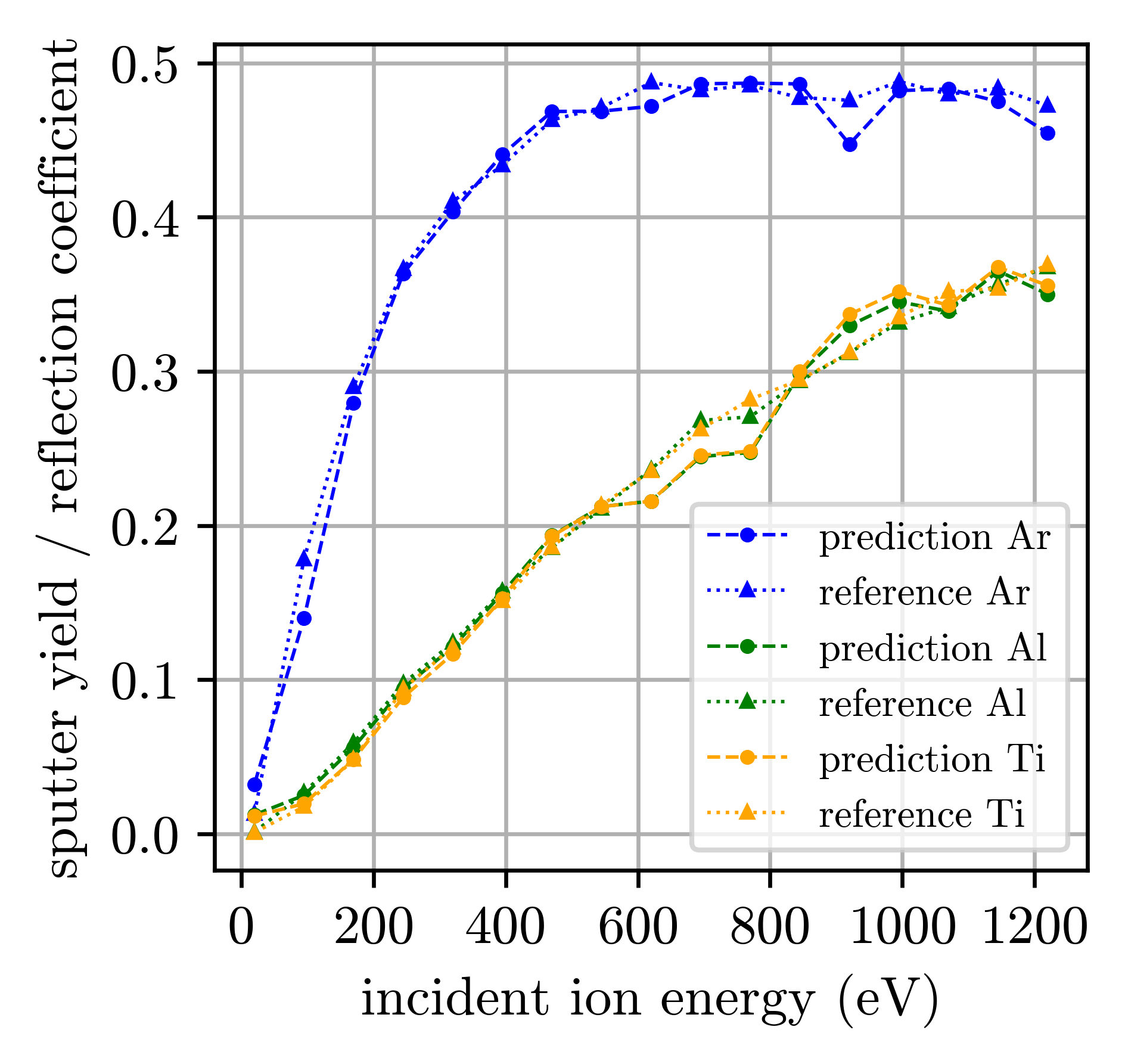}
\end{figure}

Since the respective EADs are not normalized, integrating the EADs over both energy and polar angle and dividing by the total number of incident projectiles gives the sputter yields $Y_\text{Al}$ and $Y_\text{Ti}$ as well as the reflection coefficient $r_\text{Ar}$ for Ar, respectively. Figure~\ref{fig:yield} shows the corresponding coefficients as a function of incident energy. Evidently, the predicted dependencies again appropriately capture the test data. Small deviations between the predicted and the reference results are observed for some energies. These, however, may be attributed to statistical fluctuations in the training data. A physical interpretation or a specific attribution to features of the trained ANN or to aspects of the machine learning procedure cannot be inferred. A peculiar aspect of physical reliability relates to similar sputter yields for Al and Ti: The fluxes of both metal species emitted from the target surface under continuous ion bombardment are necessarily identical on average. This is a direct consequence of mass conservation in steady state. It is instructive to observe from the Figure~\ref{fig:yield} that this criterion is similarly fulfilled for the predicted and the reference yields. The mean and maximum deviation for the predicted results are 3.9\% and 13\%, respectively. These results underline the physical relevance and reliability of the proposed procedure not only for a single interaction case, but over the complete range of incident energies previously trained.

\section{Conclusion}
\label{sec:conclusion}

This work demonstrates the successful supervised learning of statistically disturbed data sets from the solid surface model TRIDYN \cite{moller_tridyn_1984} using a multilayer perceptron ANN. For a set of input energy distributions of impinging projectile particles, the energy and angular distributions of reflected and sputtered particles are shown to be accurately predicted. The trained ANN is verified to reliably generalize, minimizing the predictions' dependence on the noise contained within the training data. Specifically, with a coefficient of determination of $R_\text{pred}^2  = 99.85\%$ the ANN corresponds more accurately to high quality reference data than the noisy training set with $R_\text{ref}^2 = 96.12\%$. This is in particular valuable in view of the limited size of the training set. This result was enabled by investigating the influence of the governing hyperparameters, which encompasses a variation of the chosen data structure as well as an investigation of the corresponding prediction times. The latter are on the order of $\Delta t_\text{pred} \approx 1$~ms and, therewith, much smaller than the approximate gas-phase simulation evaluation time per time step of $10$~ms.

From the presented results it can be inferred that our approach is a promising alternative to current methods for interfacing multi-scale models. However, apart from the determined conceptual validity, two farther-reaching aspects seem important to be addressed in the future: Firstly, the predictions' physical validity needs to be rigorously verified (e.g., distributions non-negative, flux balance fulfilled). Correspondingly, the question should be addressed whether specified aspects can be mathematically proven or inferred from the design of the network. Secondly, the ANN needs to be implemented within a comprising simulation model. In this respect, initially the ANN should be embedded as a simple plasma-surface interface model. The complexity should subsequently be increased by considering more complex chemical and molecular interactions (i.e., including reactive species and corresponding processes). This logically implies the application of more sophisticated molecular dynamics output data for training.

\section*{Acknowledgement}
The authors sincerely thank Professor Dr.-Ing.\ Thomas Mussenbrock from Brandenburg University of Technology Cottbus--Senftenberg for his advice and support. The authors thank Professor Dr.\ Wolfhard Möller from Institute of Ion Beam Physics and Materials Research, Helmholtz-Zentrum Dresden-Rossendorf (HZDR) for permission to use the TRIDYN simulation software. Financial support provided by the German Research Foundation (DFG) in the frame of the collaborative research centre TRR\,87 (SFB-TR\,87) is gratefully acknowledged.

\bibliography{references}

\end{document}